\documentclass[12pt,aps,pra,groupedaddress,amssymb,amsfonts,nofootinbib,tightenlines,showpacs]{revtex4} 
\usepackage{graphicx} 

\graphicspath{{./}{graphics/}}
\usepackage{times,color}
\definecolor{purple}{rgb}{0.625,0.125,0.9375}
\newcommand{\ignore}[1]{}
\newcommand{\draftcomment}[1]{{\color{blue}[#1]}}
\newcommand{\draftcommentG}[1]{{\color{red}[#1]}}
\newcommand{\details}[1]{{[\small #1]}}
\renewcommand{\draftcomment}[1]{}
\renewcommand{\draftcommentG}[1]{}
\renewcommand{\details}[1]{}

\definecolor{grey}{rgb}{0.8,0.8,0.8}

\ignore{
rm -R /tmp/obsexp; mkdir /tmp/obsexp
cp `texfls obsexp.log | perl -e '$a=<>; $a =~ s:/\S*(revtex4|natbib)\S*(\s|$)::g; print "$a\n"'` /tmp/obsexp
find /tmp/obsexp -type f -name '*.pdf' -exec pdftops -eps {} \; -exec rm {} \;
find /tmp/obsexp -type f -name '*.jpg' -exec perl -e '$f=shift; $g=f$; $g=~s/.jpg/.eps/; exec("convert $f $g");' {} \; -exec rm {} \;
perl -i -n -e 's/^

cp obsexp.bib /tmp/obsexp
(cd /tmp/obsexp; tar czvf obsexp.tar.gz *)
}

\ignore{
rm -R /tmp/obsexp; mkdir /tmp/obsexp
cp `texfls obsexp.log | perl -e '$a=<>; $a =~ s:/\S*(revtex4|natbib)\S*(\s|$)::g; print "$a\n"'` /tmp/obsexp
find /tmp/obsexp -type f -name '*.pdf' -exec pdftops -eps {} \; -exec rm {} \;
find /tmp/obsexp -type f -name '*.jpg' -exec perl -e '$f=shift; $g=f$; $g=~s/.jpg/.eps/; exec("convert $f $g");' {} \; -exec rm {} \;
perl -i -n -e 's/^

(cd /tmp/obsexp; tar czvf obsexp.tar.gz *)
}

\newcommand{\sysfnt}{\mathsf}

\newcommand{\ket}[1]{{|}{#1}{\rangle}}
\newcommand{\bra}[1]{{\langle}{#1}{|}}

\newcommand{\ketbra}[2]{\ket{#1}\bra{#2}}
\newcommand{\kets}[2]{{|}{#1}{\rangle}_{{}_{\!\!\scriptstyle{\sysfnt{#2}}}}}
\newcommand{\bras}[2]{{}^{\scriptstyle\sysfnt{ #2}}\!{\langle}{#1}{|}}

\newcommand{\ketbras}[3]{\kets{#1}{#3}\!\!\bras{#2}{#3}}
\newcommand{\slb}[2]{{{#1}^{({\sysfnt{#2}})}}}
\newcommand{\cntrl}[1]{{{}^c\!#1}}

\newcommand{\cQ}{{\cal Q}}

\renewcommand{\tensor}{\otimes}
\newcommand{\trace}{\mbox{tr}}

\newcommand{\one}{\mathbf{I}}

\renewcommand{\Re}{\mathrm{Re}}
\renewcommand{\Im}{\mathrm{Im}}

\begin{document}
\title{Optimal Quantum Measurements of Expectation Values of Observables}
\author{Emanuel Knill}
\email[]{knill@boulder.nist.gov}
\affiliation{Mathematical and Computational Sciences Division, National
Institute of Standards and Technology, Boulder CO 80305}
\author{Gerardo Ortiz}
\email[]{g\underline{\ }ortiz@lanl.gov, ortizg@indiana.edu}
\affiliation{Theoretical Division, Los Alamos National Laboratory, Los 
Alamos, NM 87545 }
\affiliation{Department of Physics, Indiana University, Bloomington,
IN 47405}
\author{Rolando D. Somma}
\email[]{somma@lanl.gov}
\affiliation{Physics Division, Los Alamos National Laboratory, Los Alamos, NM 87545 }

\date{\today}
\begin{abstract}
Experimental characterizations of a quantum system involve the
measurement of expectation values of observables for a preparable
state $\ket{\psi}$ of the quantum system. Such expectation values can
be measured by repeatedly preparing $\ket{\psi}$ and coupling the
system to an apparatus. For this method, the precision of the measured
value scales as ${1\over\sqrt{N}}$ for $N$ repetitions of the
experiment.  For the problem of estimating the parameter $\phi$ in an
evolution $e^{-i\phi H}$, it is possible to achieve precision $1\over
N$ (the quantum metrology limit, see~\cite{giovannetti:qc2006a})
provided that sufficient information about $H$ and its spectrum is
available. We consider the more general problem of estimating
expectations of operators $A$ with minimal prior knowledge of $A$. We
give explicit algorithms that approach precision $1\over N$ given
a bound on the eigenvalues of $A$ or on their tail distribution.  These
algorithms are particularly useful for simulating quantum systems on
quantum computers because they enable efficient measurement of
observables and correlation functions.  Our algorithms are based on a
method for efficiently measuring the complex overlap of $\ket{\psi}$
and $U\ket{\psi}$, where $U$ is an implementable unitary operator. We
explicitly consider the issue of confidence levels in measuring observables
and overlaps and show that, as expected, confidence levels can be improved
exponentially with linear overhead. We further show that the
algorithms given here can typically be parallelized with minimal
increase in resource usage.
\end{abstract}

\maketitle

\section{Introduction}

Uncertainty relations such as Heisenberg's set fundamental physical
limits on the achievable precision when we extract information from a
physical system.  The goal of quantum metrology is to measure
properties of states of quantum systems as precisely as possible given
available resources. Typically, these properties are determined by
experiments that involve repeated preparation of a quantum system in a
state $\rho$ followed by a measurement. The property is derived from
the measurement outcomes.  Because the repetitions are statistically
independent, the precision with which the property is obtained scales
as ${1\over\sqrt{N}}$, where $N$ is the number of preparations
performed.  This is known as the standard quantum limit or the
shot-noise limit, and it is associated with a purely classical
statistical analysis of errors.  It has been shown that in many cases
of interest, the precision can be improved to ${1\over N}$ by using
the same resources but with initial states entangled over multiple
instances of the quantum system, or by preserving quantum coherence
from one experiment to the next.  It is known that it is usually not
possible to attain a precision that scales better than ${1\over
N}$. (See~\cite{giovannetti:qc2004a} for a review of quantum-enhanced
measurements.)  A setting where this limit can be achieved is the
parameter estimation problem, where the property is given by the
parameter $\phi$ in an evolution $e^{-i\phi H}$ for a known
Hamiltonian $H$~\cite{giovannetti:qc2006a}, which captures some common
measurement problems. The standard method for determining $\phi$
requires the ability to apply $e^{-i\phi H}$ and to prepare and
measure an eigenstate of $H$ with known eigenvalue. If it is not
possible to prepare such an eigenstate or if we wish to determine
expectations with respect to arbitrary states, this method fails.
Here we are interested in the more general and physically important
expectation estimation problem, where the property to be determined is
an expectation $\langle A \rangle=\trace(A\rho)$ of an observable
(Hermitian operator) or unitary $A$, for a possibly mixed state
$\rho$.  Both $A$ and $\rho$ are assumed to be experimentally
sufficiently controllable, but other than a bound on the eigenvalues
of $A$ or their tail distribution, no other properties of $A$ or
$\rho$ need to be known.  In particular, we need not be able to
prepare eigenstates of $A$ or know the spectrum of $A$.  The parameter
estimation problem is a special instance of the expectation estimation
problem. Parameter estimation reduces to the problem of determining
$\trace(e^{-i\phi H}\ketbra{\psi}{\psi})$ for $\ket{\psi}$ an
eigenstate of $H$ with non-zero eigenvalue.  We show that for solving
the expectation estimation problem, precision scalings of ${1\over
N^{1-\alpha}}$ for arbitrarily small $\alpha>0$ can be achieved with
sequential algorithms, and the algorithms can be parallelized with
minimal additional resources.

\begin{figure}
\includegraphics[width=3.3in]{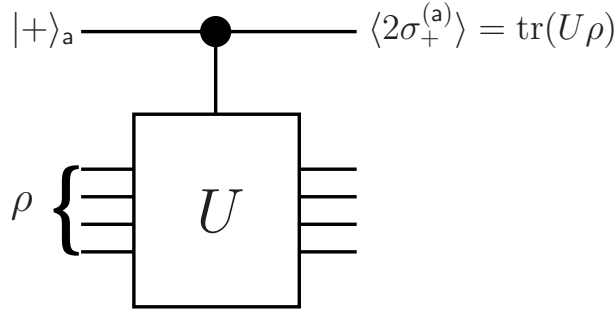}
\caption{Quantum network for the one-ancilla algorithm to measure
$\langle U \rangle=\trace(U\rho)$ with $\ket{+}_{\sf a}=(\ket{0}_{\sf
a}+\ket{1}_{\sf a})/\sqrt{2}$ in the logical basis. The desired
expectation is given by $\trace(U\rho)=\langle 2\sigma_+^{\sf a}\rangle
= \langle\sigma_x^{({\sf a})}\rangle+i\langle\sigma_y^{({\sf
a})}\rangle$, where $\langle\sigma_x^{({\sf a})}\rangle$ and
$\langle\sigma_y^{({\sf a})}\rangle$ are the expectations of the Pauli
matrices $\sigma_x^{({\sf a})}$ and $\sigma_y^{({\sf a})}$ for the
final state, which are estimated by repeating the experiment and
measuring either $\sigma_x^{({\sf a})}$ or $\sigma_y^{({\sf a})}$ on
the control (ancilla) qubit. Because these measurements have $\pm 1$
as possible outcomes, their statistics are determined by the binomial
distribution.}
\label{fig:onealg}
\end{figure}

Our motivation for this work is the setting of quantum physics
simulations on quantum computers. This is one of the most promising
applications of quantum computing~\cite{feynman:qc1982a} and enables a
potentially exponential speedup for the correlation function
evaluation problem~\cite{terhal:qc1998a,ortiz:qc2000a,somma:qc2003a}.
The measurement of these correlation functions reduces to the
measurement of the expectation of an operator for one or more states.
Because the measurement takes place within a scalable quantum
computer, the operators and states are manipulatable via arbitrarily
low-error quantum gates.  The quantum computational methods that have
been described for the determination of these expectations have order
${1\over\sqrt{N}}$ precision. An example is the one-ancilla
algorithm for measuring $\langle U \rangle=\trace(U\rho)$ for unitary
$U$ described in~\cite{ortiz:qc2000a,somma:qc2001a,miquel:qc2001a},
which applies $U$ conditional on an ancilla ${\sf a}$ prepared in a
superposition state (Fig.~\ref{fig:onealg}). Improving the precision
without special knowledge of the operator or state requires more
sophisticated algorithms.

Here we give quantum algorithms based on phase and amplitude
estimation~\cite{cleve:qc1997b,brassard:qc2000a} to improve the
resource requirements to achieve a given precision. We begin by giving
an ``overlap estimation'' algorithm (OEA) for determining the
amplitude and phase of $\trace(U\rho)$ for $U$ unitary. We assume that
quantum procedures for preparing $\rho$ from a standard initial state
and for applying $U$ are known and that it is possible to reverse
these procedures.  We determine the number of times $N$ that these
procedures are used to achieve a goal precision $p$ and show that $N$
is of order $1/p$.  To determine $\trace(A\rho)$ for observables $A$
not expressible as a small sum of unitary operators, we assume that it
is possible to evolve under $A$. This means that we can apply
$e^{-iAt}$ for positive times $t$. The OEA can be used to obtain
$\trace(A\rho)t \approx i(\trace(e^{-iAt}\rho)-1)$ for small $t$. The
problem of how to measure $\trace(A\rho)$ with precision $p$ requires
determining $\trace(e^{-iAt}\rho)$ with precision better than $pt$ and
choosing $t$ small enough that the error in the approximation does not
dominate.  We solve this problem by means of an ``expectation
estimation'' algorithm (EEA) with minimal additional knowledge on the
eigenvalue distribution of $A$. For this situation, the relevant
resources are not only the number $N$ of uses of $e^{-iAt}$ and of the
state preparation algorithm, but also the total time $T$ of evolution
under $A$.  We show that to achieve a goal precision $p$, $N$ and $T$
are of order $1/(p^{1+\alpha})$ and $1/p$, respectively, with
$\alpha>0$ arbitrarily small. The term $\alpha$ in the resource bound
is due partly to the tail distribution of the eigenvalues of $A$ with
respect to $\rho$. When it is known that $\rho$ is an eigenstate of
$A$, so the distribution is a delta function, $\alpha=0$. This applies
to the parameter estimation problem. In the case where $A$ is
unbounded, $\alpha$ is still arbitrarily small if the tail
distribution is exponentially decaying. But if only small moments of
$A$ can be bounded, in which case the best bound on the tail
distribution decays polynomially, $\alpha$ becomes finite.

It is important to properly define the meaning of the term
``precision''.  Here, when we say that we are measuring $\trace(A\rho)$ with
precision $p$, we mean that the probability that the measured value
$a_{\mathrm{meas}}$ is within $p$ of $\trace(A\rho)$ is bounded below
by a constant $c>0$. In other words, the ``confidence level'' that
$a_{\mathrm{meas}}-p \leq \trace(A\rho) \leq a_{\mathrm{meas}}+p$ is
at least $c$. Thus $a_{\mathrm{meas}}\pm p$ defines ``confidence
bounds'' of the measurement for confidence level $c$. One
interpretation of confidence levels is that if the measurement is
independently repeated, the fraction of times the measured value is
within the confidence bound is at least the confidence level.  For
measurement values $a_{\mathrm{meas}}$ that have an (approximately)
gaussian distribution, it is conventional to use $c=0.68$ to identify
the precision $p$ with the standard deviation. In this case, the
confidence level that the measurement outcome is within $xp$ can be
bounded by $\mathrm{erf}(x/\sqrt{2})$, where $\mathrm{erf}(y)$ is the
error function, $\mathrm{erf}(x/\sqrt{2}) \geq 1- e^{-x^2/2}$. This
bound is often too optimistic, which is one reason to specify
confidence levels explicitly. This becomes particularly important in
our use of the ``phase estimation" algorithm (PEA), whose standard
version~\cite{cleve:qc1997b} has confidence levels that converge
slowly toward $1$ with $x$.  Because of these issues, our algorithms
are stated so that they solve the problem of determining
$\trace(A\rho)$ with precision $p$ and confidence level $c$, where $p$
and $c$ are specified at the beginning. This requires that the
resource usage be parameterized by both $p$ and $c$, and we show that
the resource usage grows by a factor of order $|\log(1-c)|$ to achieve
high confidence level $c$.

An important problem in measuring properties of quantum systems is how
well the measurement can be parallelized with few additional
resources. The goal of parallelizing is to minimize the time for the
measurement by using more parallel resources. Ideally, the time for
the measurement is independent of the problem.  Typically we are
satisfied if the time grows at most logarithmically.  It is well known
that for the parameter estimation problem, one can readily parallelize
the measurement by exploiting entanglement in state
preparation~\cite{bollinger:qc1996a}.  That this is still possible for
the OEA and EEA given here is not obvious. In fact, we show that
there are cases where parallelization either involves a loss of
precision or requires additional resources.  However, the entanglement
method for parallelizing measurements works for expectation estimation
and for overlap estimation when $|\trace(U\rho)|$ is not close to $1$.

\details{Note: Text using a reduced size font in square brackets gives
details of arguments and calculations.  We include them for
completeness. They are generally straightforward but tedious and do not
significantly contribute to the discussion.}
\draftcomment{Recommendation: Remove the reduced-font
details as well as the draft comments (in blue) before
publication by uncommenting the appropriate LaTeX command
redefinitions.}

\section{Overlap Estimation}
\label{sect:unitexp}

Let $U$ be a unitary operator and $\rho$ a state of quantum system
$\sysfnt{S}$. We assume that we can prepare $\rho$ and apply $U$ to any
quantum system $\sysfnt{S'}$ that is equivalent to $\sysfnt{S}$.  Both
the preparation procedure and $U$ must be reversible.  In addition, we
require that the quantum systems are sufficiently controllable and that
$U$ can be applied conditionally (see below). We use labels to clarify
which quantum system is involved. Thus, $\slb{\rho}{S'}$ is the state
$\rho$ of system $\sysfnt{S'}$ and $\slb{U}{S'}$ is $U$ acting on
system $\sysfnt{S'}$.  This allows us to prepare $\rho$ and apply $U$
in parallel on multiple quantum systems. 

When we say that we can prepare $\rho$, we mean that we can do this
fully coherently. That is, we have access to a unitary operator
$V^{(\sysfnt{SE})}$ that can be applied to a standard initial state
$\kets{0}{SE}$ of $\sysfnt{S}$ and an ancillary system $\sysfnt{E}$
(environment) such that $\slb{\rho}{S}= \trace_\sysfnt{E}
(V^{(\sysfnt{SE})}\ketbras{0}{0}{SE}(V^{(\sysfnt{SE})})^\dagger)$. The
state $V^{(\sysfnt{SE})}\kets{0}{SE}$ is a so-called purification of
$\slb{\rho}{S}$.  For our purposes and without loss of generality, we
can assume that $\rho$ is pure by merging systems $\sysfnt{E}$ and
$\sysfnt{S}$ and letting unitaries act on the merged system. With this
simplification we can write $\rho=\ketbra{\psi}{\psi} =
V\ketbra{0}{0}V^\dagger$ and use $\sysfnt{S},\sysfnt{S'},\ldots$ to
refer to equivalent merged systems. The goal of the OEA is now to
estimate the overlap $\bra{\psi}U\ket{\psi}$ of $\ket{\psi}$ with
$U\ket{\psi}$.

The OEA and EEA require that $\sysfnt{S}$ is sufficiently
controllable. In particular, we require that it is possible to couple
$\sysfnt{S}$ to ancilla qubits and to implement conditional selective
sign changes of $\kets{0}{S}$.  Let $\slb{P_0}{S} =
\slb{\one}{S}-2\ketbras{0}{0}{S}$ be the selective sign change of 
$\kets{0}{S}$, with $\slb{\one}{S}$ the identity (or no-action)
operator. If an ancilla (control) qubit is labeled $\sysfnt{a}$, an
instance of the conditional selective sign change is defined by
\begin{equation}
\slb{\cntrl{P_0}}{aS} = 
\ketbras{0}{0}{a}\slb{\one}{S}+\ketbras{1}{1}{a}\slb{P_0}{S} .
\end{equation}
If $\sysfnt{S}$ consists of qubits and $\kets{0}{S}$ is the usual
starting state with all qubits in logical state $\ket{0}$, then this is
essentially a many-controlled sign flip and has efficient
implementations~\cite{barenco:qc1995a}. 

As mentioned above, for the OEA we require
that $U$ can be applied conditionally. This means that the unitary
operator
\begin{equation}
\slb{\cntrl{U}}{aS} = 
\ketbras{0}{0}{a}\slb{\one}{S}+\ketbras{1}{1}{a}\slb{U}{S}
\end{equation}
is available for use. When $U$ is associated with an evolution
simulated on a quantum computer, this is no problem since all quantum
gates are readily ``conditionalized''~\cite{barenco:qc1995a}.
Nevertheless, we note that $\cntrl{U}$ is not required if only the amplitude
$|\bra{\psi}U\ket{\psi}|$ of $\bra{\psi}U\ket{\psi}$ is needed.

The ``amplitude estimation" algorithm (AEA)~\cite{brassard:qc2000a}
can almost immediately be applied to obtain $|\bra{\psi}U\ket{\psi}|$.
To accomplish our goals we need to adapt it for arbitrarily prepared
states and use a version that avoids the complexities of the full
quantum Fourier transform~\cite{shor:qc1995a}. Before we describe and
analyze the version of the AEA needed here, we show how the OEA uses
it to estimate the phase and amplitude of $\bra{\psi}U\ket{\psi}$.
Let $\mathrm{AE}(U,\ket{\psi},p)$ be the estimate of
$|\bra{\psi}U\ket{\psi}|$ obtained by the AEA for goal precision $p$.
(We specify the meaning of the precision parameter below.)

\begin{figure}
     \begin{center}
      \setlength{\unitlength}{1cm}
       \resizebox*{6in}{!}{
       \begin{picture}(22,10)(-8,6)            

\put(-2.0,6){\includegraphics[scale=0.75]{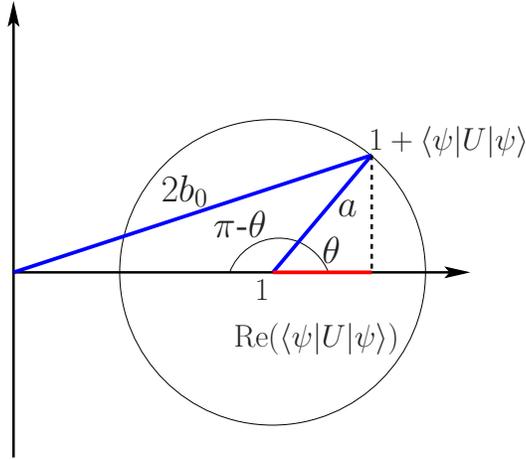}}
        \put(1.0,11.0){\LARGE $2b_0$}
         \put(4.4,10.7){\LARGE $a$}
         \put(2.0,10.3){\LARGE $\pi$-$\theta$}
         \put(4.1,9.8){\LARGE $\theta$}
        \put(2.8,9.1){\Large $1$}
        \put(2.4,8.2){\Large ${\rm Re}(\bra{\psi} U \ket{\psi})$}
        \put(5,12.0){\Large $1+\bra{\psi} U \ket{\psi}$}
      \thicklines
       \end{picture}}
      \end{center}
\caption{Geometrical construction for computing
$\Re(\bra{\psi}U\ket{\psi})$ from $a=|\bra{\psi}U\ket{\psi}|$ and
$2b_0=|(1+\bra{\psi}U\ket{\psi})|$.  According to the law of cosines,
$(2b_0)^2=a^2+1+2a\cos(\theta)$, and we have
$\Re(\bra{\psi}U\ket{\psi})=a\cos(\theta)=
((2b_0)^2-a^2-1)/2$.}
\label{fig:rey}
\end{figure}

\begin{figure}
     \begin{center}
      \setlength{\unitlength}{1cm}
       \resizebox*{5.5in}{!}{
       \begin{picture}(22,16)(-4,4)            
        \put(-3.45,4.3){\includegraphics[scale=0.75]{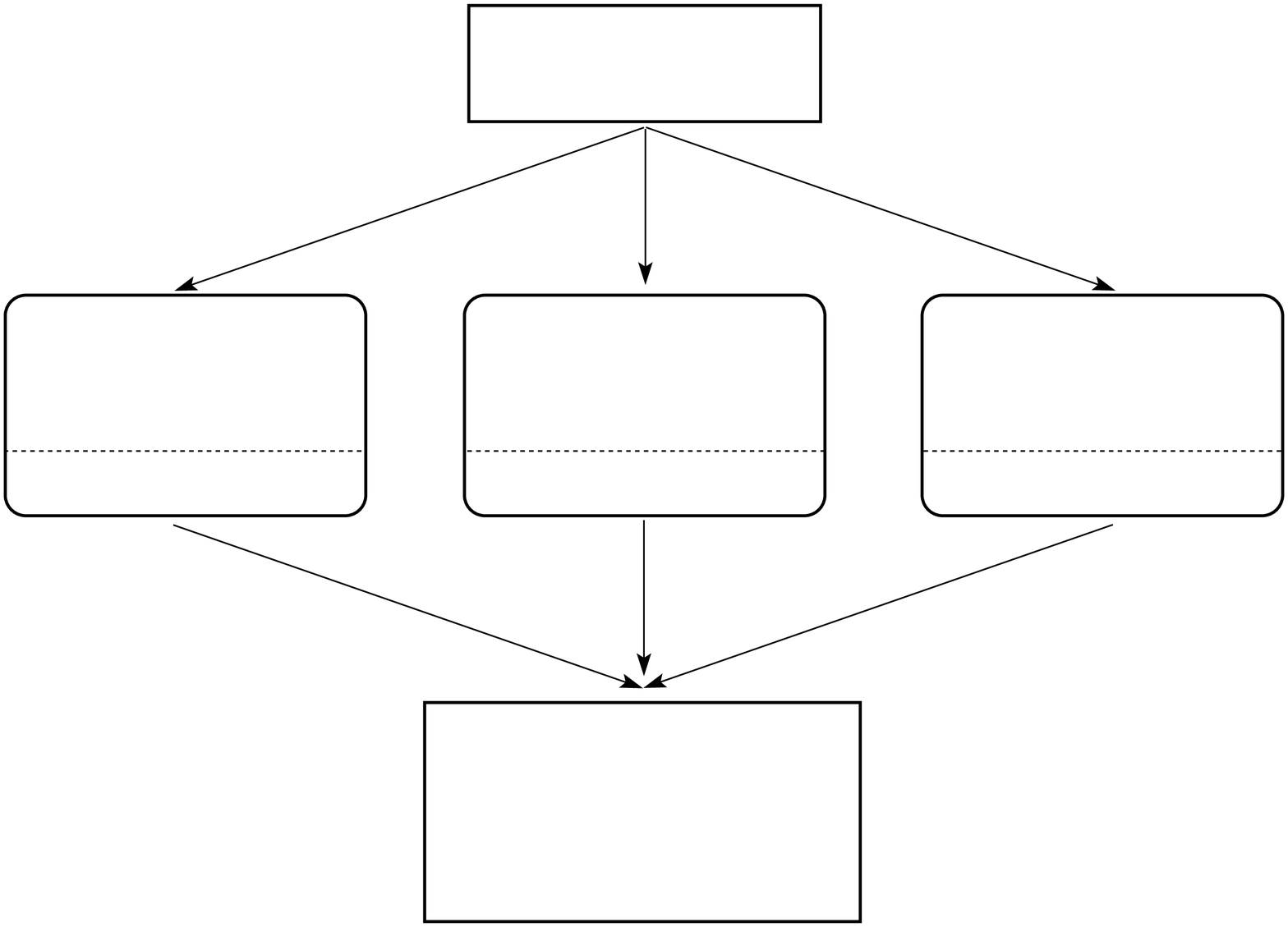}}
        \put(-2.5,13.3){\Large $a=$AE$(U, |\psi \rangle, \frac{p}{4})$}
        \put(-2.2,11.4){\Large $a\approx|\langle \psi | U | \psi \rangle|$}
        \put(4.07,13.3){\Large $b_0\!\!=$AE$(\cntrl{U}^{(\sf aS)}\!, \!|\!\!+\!\psi
        \rangle_{\!\sf aS},\! \frac{p}{16})$}
        \put(4.3,11.4){\Large $b_0\approx|1+\langle \psi | U | \psi \rangle|/2$}
        \put(11.6,13.3){\Large $b_{\pi/2}\!=$AE$(\tilde{U}, \!|\!\!+\!\psi
        \rangle_{\! \sf aS},\! \frac{p}{16})$}
         \put(12.1,12.3){\Large $\tilde{U}=e^{i\sigma_z^{(\sf
         a)}\pi/4}cU^{(\sf aS)}$}
       \put(11.5,11.4){\Large $b_{\pi/2}\!\approx\!|1-\!i\langle \psi | U | \psi \rangle|/2$}
        \put(-2.5,15.5){\LARGE \color{red} DO:}
        \put(-2.5,8.7){\LARGE \color{red} ESTIMATE $y$:}
        \put(-2.5,8.0){\Large \color{red} (with precision $p$)}
        \put(5.9,18.5){\LARGE \color{red} INPUT}
        \put(5.7,17.5){\LARGE $U, |\psi \rangle, p$}
        \put(3.6,7.00){\Large ${\rm Re}(y)=(4b_0^2-a^2-1)/2$}
        \put(3.6,6.){\Large ${\rm Im}(y)=(4b_{\pi/2}^2-a^2-1)/2$}
        \put(3.5,5.){\Large $y\!=\!\langle \psi | U | \psi \rangle\!=\!{\rm
        OE}(U, |\psi \rangle, p)$}
      \thicklines
       \end{picture}}
      \end{center}
\caption{OEA flowchart. An estimate of the overlap $\langle \psi | U | \psi
\rangle$ is obtained.  The algorithm requires three state preparations
and calls the AEA three times. The amplitude of the returned value
shown in the flowchart may need to be adjusted according to the value
of $a$ to optimize the precision. For details see the text.}
\label{ov-est}
\end{figure}

\begin{list}{}{}
\item[\textbf{Overlap estimation algorithm:}] Given are $U$,
$\ket{\psi}$ (in terms of a preparation unitary
$V:\ket{0}\mapsto\ket{\psi}$) and the goal precision $p$. An estimate
of $\bra{\psi}U\ket{\psi}$ is to be returned.
\item[1.] Obtain $a=\mathrm{AE}(U,\ket{\psi},p/4)$, 
   so that $a$ is an estimate of $|\bra{\psi}U\ket{\psi}|$ with
   precision $p/4$.
\item[2.] Obtain $b_0=\mathrm{AE}(\slb{\cntrl{U}}{aS},\kets{+\psi}{aS}=\kets{+}{a}\ket{\psi},p/16)$. 
          
	  Note that $\bras{+\psi}{aS}\slb{\cntrl{U}}{aS}\kets{+\psi}{aS}=(1+\bra{\psi}U\ket{\psi})/2$.
\item[3.] Obtain
$b_{\pi/2}=\mathrm{AE}(e^{i\slb{\sigma_z}{a}\pi/4}\slb{\cntrl{U}}{aS},\kets{+\psi}{aS},p/16)$.

Note that
$\bras{+\psi}{aS}e^{i\slb{\sigma_z}{a}\pi/4}\slb{\cntrl{U}}{aS}\kets{+\psi}{aS}=
e^{i\pi/4}(1-i\bra{\psi}U\ket{\psi})/2$.
\item[4.] Estimate the phase $\theta$  of $\bra{\psi}U\ket{\psi}$ by
computing the argument of the complex number $y$ defined by
\begin{eqnarray}
\Re(y) &=& (4b_0^2-a^2-1)/2,\nonumber\\
\Im(y) &=& (4b_{\pi/2}^2-a^2-1)/2.
\label{eqn:reim}
\end{eqnarray}
If $a$, $b_0$ and $b_{\pi/2}$ were the exact values of the amplitudes
estimated by the three instances of the AEA, then we would have
$y=\bra{\psi}U\ket{\psi}$. For example, the formula for $\Re(y)$ may be
obtained by geometrical reasoning, as shown in Fig.~\ref{fig:rey}.
\item[5.] Estimate $\bra{\psi}U\ket{\psi}$ as $e^{i\theta}a$. The
reason for not using $y$ directly is that  if the overlap has amplitude
near $1$, then the error in the amplitude of $y$ can be substantially larger
than the error in $a$. (This is because of the way we estimate $y$ using a 
PEA; see below.)
\end{list}
We define $\mathrm{OE}(U,\ket{\psi},p)$ to be the value returned by
the OEA. A flowchart for the algorithm is depicted in Fig.~\ref{ov-est}.

\begin{figure}
    \begin{center}
      \setlength{\fboxrule}{0pt}\setlength{\fboxsep}{0pt}
      \setlength{\unitlength}{1cm}
       \resizebox*{4.3in}{!}{
       \begin{picture}(22,15)(-5,4)            
        \put(0,4.3){\includegraphics[scale=0.75]{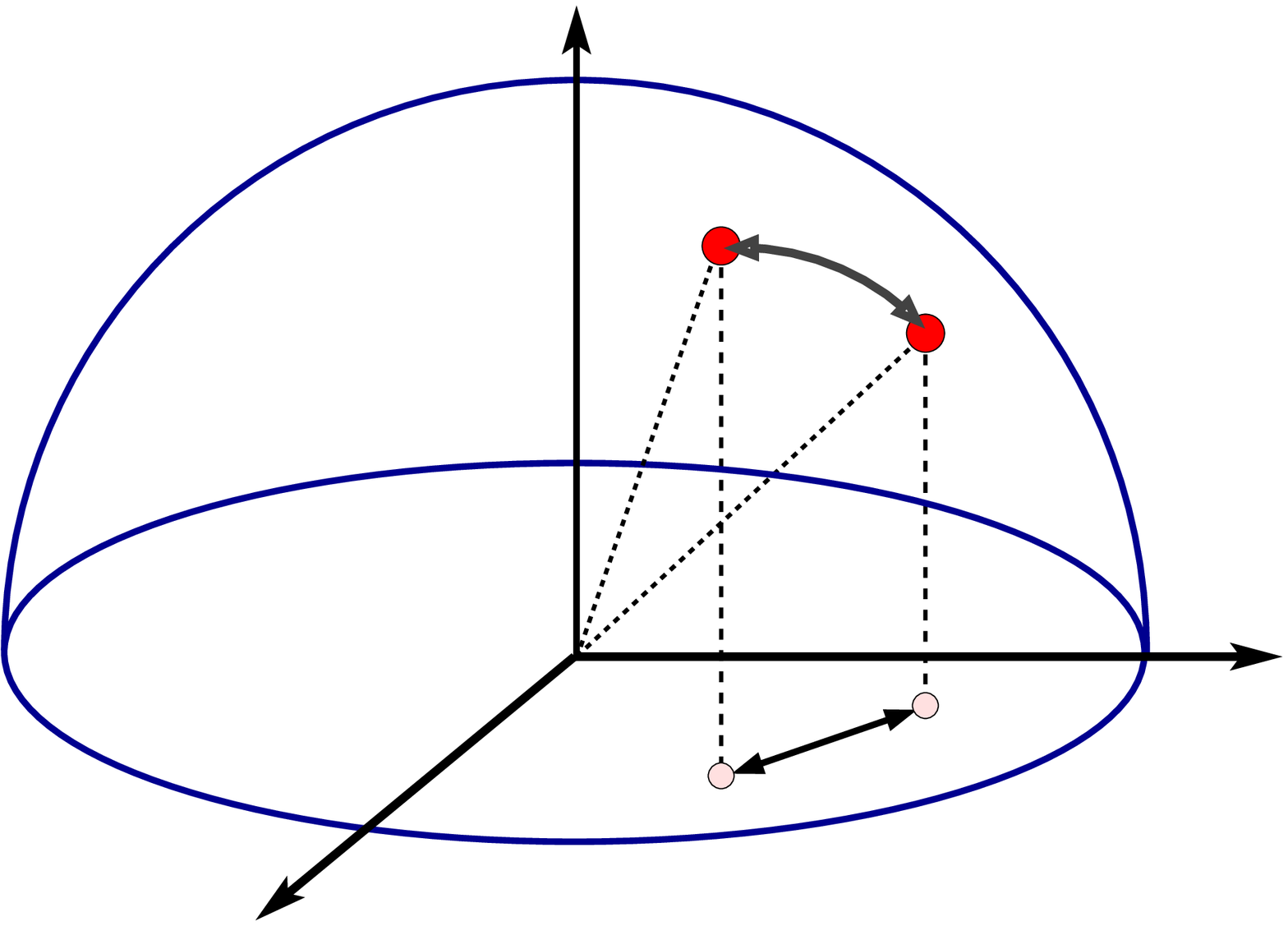}}
        \put(4,12.4){\fcolorbox[rgb]{1,1,1}{1,1,1}{\Huge $h^{-1}(\langle \psi | U | \psi \rangle)$}}
        \put(10.70,11.2){\fcolorbox[rgb]{1,1,1}{1,1,1}{\Huge $h^{-1}(o)$}}
        \put(8.90,10.9){\Huge $\delta$}
        \put(8.90,6.5){\Huge $\delta'$}
        \put(6.8,14.4){\Huge $x_3$}
        \put(4.95,5.60){\Huge $\langle \psi | U | \psi \rangle$}
        \put(10.7,6.8){\Huge $o$}
        \put(13.5,7.8){\Huge $x_2$}
        \put(2.0,4.8){\Huge $x_1$}
      \thicklines
       \end{picture}}
      \end{center}
\caption{Visualization of the parameterization of the overlap in terms
of points on the upper hemisphere of a unit sphere.  The function $h$
is defined by $h(x_1,x_2,x_3)=x_1+i x_2$.  Note that for overlaps
$|\bra{\psi}U\ket{\psi}|$ approaching $1$ and small $\delta$,
$\delta'$ approaches $\delta^2/2\ll\delta$.}
\label{lift}
\end{figure}

When $a=|\bra{\psi}U\ket{\psi}|$ is close to $1$, the absolute
precision with which $a$ is obtained is as much as quadratically
better for the same resources.  To avoid this nonuniformity of the
precision to resource relationship, we define the precision $\delta$
of an overlap by means of a parameterization of
$\bra{\psi}U\ket{\psi}$ using the points $(x_1,x_2,x_3)$ on the upper
hemisphere of the surface of a unit sphere in three dimensions.  For
this purpose, define $h(x_1,x_2,x_3) =x_1+ix_2$ for
$x_1^2+x_2^2+x_3^2=1$ and $x_3\geq 0$. Define the distance between
$(x_1,x_2,x_3)$ and $(x_1',x_2',x_3')$ to be the angular distance
along a great circle.  The precision of the value $o$ returned by the
OEA is determined by the distance $\delta$ between the liftings $h^{-1}(o)$ and
$h^{-1}(\bra{\psi}U\ket{\psi})$ (see Fig.~\ref{lift}). We define the
precision of the value returned by the AEA similarly, by restricting
the parametrization to the positive reals.  The precision parameters
with which the AEA is called in the OEA are chosen so that the returned
overlap has precision $\delta\leq p$ with respect to our
parametrization (see Note~\cite{note1}).

\begin{figure}
\includegraphics*[width=3in]{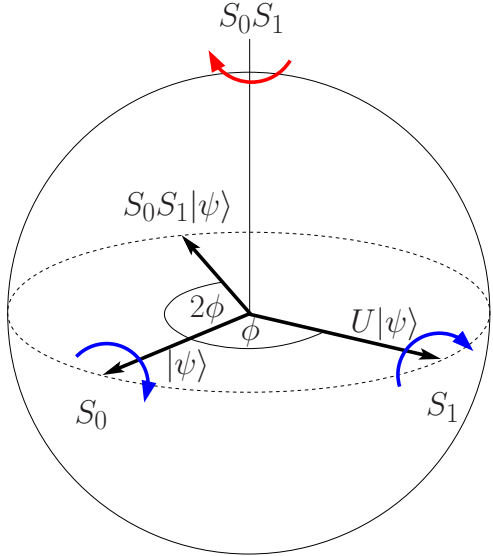}
\caption{Bloch sphere picture of the rotations induced on the
subspace spanned by $\ket{\psi}$ and $U\ket{\psi}$ by the operators
$S_0$ and $S_1$.}
\label{amp-est}
\end{figure}

The AEA is based on a trick for converting amplitude into phase
information, so that an efficient PEA can be applied. Let
$\ket{\psi_0}=\ket{\psi}$ and $\ket{\psi_1}=U\ket{\psi}$.  Let
$S_0=\one-2\ketbra{\psi_0}{\psi_0} = VP_0V^\dagger$ be the selective
sign change of $\ket{\psi_0}$ and $S_1=\one-2\ketbra{\psi_1}{\psi_1} =
UVP_0V^\dagger U^\dagger$ the selective sign change of $\ket{\psi_1}$.
The composition $S=S_0S_1$ is a unitary operator that rotates
$\ket{\psi_0}$ toward $\ket{\psi_1}$ in the two-dimensional subspace
$\cQ$ spanned by $\ket{\psi_0}$ and $\ket{\psi_1}$. The rotation is by
a Bloch-sphere angle of $2\phi=4\arccos(| \langle\psi_0|\psi_1\rangle
|)$.  Thus, the eigenvalues of $S$ in $\cQ$ are $e^{\pm i\phi}$.  The
Bloch sphere picture of the states and the rotation are shown in
Fig.~\ref{amp-est}. When $| \langle\psi_0|\psi_1\rangle
|=|\bra{\psi}U\ket{\psi}|=1$, $S$ is the identity operator. The PEA
for $S$ with initial state $\ket{\psi_0}$ determines the phase $\phi$
of one of these eigenvalues, where each of the signs has equal
probability of being returned. The overlap $|\bra{\psi}U\ket{\psi}|$
is obtained from $\phi$ by the formula $|\bra{\psi}U\ket{\psi}| =
\cos(\phi/2)$.  The PEA requires use of the conditional $S$ operator,
$\cntrl{S}$. As defined, this needs to be decomposed into a product of
$\cntrl{P_0}$, $\cntrl{U}$ and $\cntrl{V}$. A significant simplification is to not
condition $U$ and $V$ and to write $\cntrl{S}=V\cntrl{P_0}V^\dagger UV\cntrl{P_0}V^\dagger
U^\dagger$. This works because if the controlling qubit is in state
$\ket{0}$, all the $U$'s and $V$'s are canceled by matching
$U^\dagger$'s and $V^\dagger$'s~\cite{somma:qc2001a}.

Let $\mathrm{PE}(W,\ket{\psi'},p)$ be a phase returned by the PEA for
unitary operator $W$ and initial state $\ket{\psi'}$ with precision
goal $p$. The AEA may be summarized as follows.

\begin{list}{}{}
\item[\textbf{Amplitude estimation algorithm:}] Given are $U$,
$\ket{\psi}$ (in terms of a preparation unitary
$V:\ket{0}\mapsto\ket{\psi}$) and the goal precision $p$. An estimate
of $|\bra{\psi}U\ket{\psi}|$ is to be returned.
\item[1.] Let $\phi=\mathrm{PE}(S,\ket{\psi},2p)$ with 
$S=S_0S_1=VP_0V^\dagger UVP_0V^\dagger U^\dagger$.
\item[2.] Estimate  $|\bra{\psi}U\ket{\psi}|$ as $|\cos(\phi/2)|$.
\end{list}

The precision parameter for the PEA has the conventional interpretation
(modulo $2\pi$). Because $\arccos(|\bra{\psi}U\ket{\psi}|)$ is the
angle along the semicircle in the parametrization of the overlap
defined above, the precision $2p$ of the value returned by the PEA
translates directly to the desired precision in the value to be
returned by the AEA.

The PEA~\cite{cleve:qc1997b} for a unitary operator $W$ and initial
state $\ket{\psi'}$ returns an estimate of the phase $\phi$
(``eigenphase'') of an eigenvalue $e^{i\phi}$ of $W$, where the
probability of $\phi$ is given by the probability amplitude of
$\ket{\psi'}$ in the $e^{i\phi}$-eigenspace of $W$. In the limit of
perfect precision, it acts as a von Neumann measurement of $W$ on
state $\ket{\psi'}$ in the sense that the final state is projected
onto the $e^{i\phi}$-eigenspace of $W$. For finite precision, the
eigenspaces may be decohered and the projection is incomplete, unless
there are no other eigenvalues within the precision bound. The error
in the projection is related to the confidence level with which the
precision bound holds.

The original PEA is based on the binary quantum Fourier transform
\cite{shor:qc1995a}.  It determines an eigenphase $\phi$ with
precision $1\over 2^{n}$ with $2^{n}-1$ uses of the conditional
$\cntrl{W}$ operator to obtain a phase kickback to ancilla qubits.
The original PEA begins by preparing $n$ qubits labeled
$\sysfnt{1}\ldots\sysfnt{n}$ in state $\kets{+}{1}\ldots\kets{+}{n}$
and system $\sysfnt{S}$ in state $\kets{\psi'}{S}$.  Next, for each
$m=1,\ldots,n$, $\cntrl{W}$ is applied from qubit $\sysfnt{m}$ to
system $\sysfnt{S}$ $2^{m-1}$ times. The binary quantum Fourier
transform is applied to the $n$ qubits, and the qubits are measured in
the logical basis $\ket{0},\ket{1}$.  The measurement outcomes give
the first $n$ digits of the binary representation of
$\phi/(2\pi)+\epsilon/2^n$, where $|\epsilon| < 1/2$ with probability
at least $0.405$~\cite{cleve:qc1997b}.

The PEA as outlined in the previous paragraph makes suboptimal use of
quantum resources. We prefer a one-qubit version of the algorithm based
on the measured quantum Fourier transform~\cite{griffiths:qc1995a} that
has been experimentally implemented on an ion trap quantum
computer~\cite{chiaverini:qc2005a}. An advantage of this approach is
that it does not require understanding the quantum Fourier transform
and is readily related to more conventional approaches for measuring
phases. To understand how the algorithm given below works, note
that the eigenstates of $W$ are invariant under $\cntrl{W}$. The only
interaction with $\sysfnt{S}$ is via uses of $\cntrl{W}$. Therefore, without
loss of generality, we can assume that $\sysfnt{S}$ is initially
projected to an $e^{i\phi}$-eigenstate of $W$ with $0\leq \phi <
2\pi$.  The bits of an approximation of $\phi/(2\pi)$ are determined
one by one, starting with the least significant one that we wish to
learn.  Given $n$, let $[.b_1\ldots b_n]_2=\sum_{i=1}^n b_i/2^i$ (with
$b_i=0,1$) be a best  $n$-digit binary approximation to $\phi/(2\pi)$,
where the notation $[x]_2$ is used to convert a sequence of binary
digits $x$ to the number that it represents.  Write $\epsilon=
(\phi/(2\pi)-[.b_1\ldots b_n]_2)2^n$.

\begin{figure}
\includegraphics*[width=5.3in]{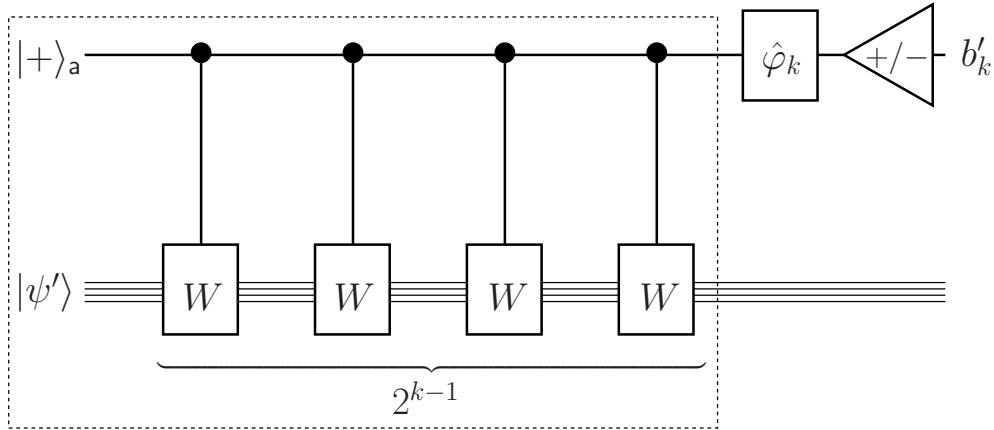}
\caption{Step 2 of the PEA to estimate bit $k$ of the eigenphase,
where $k=3$. The phase $\hat\varphi_k$ is computed according to
previously obtained information about the eigenphase. By applying it
before the measurement, the probability of obtaining the optimal value
for bit $k$ is maximized.  The measurement is denoted by the triangle
pointing left with $+/-$ inside and is a measurement in the
$\ket{+}/\ket{-}$ basis.  The outlined part of the network will be
parallelized in Sect.~\ref{sect:parallel}.}
\label{phase-est}
\end{figure}

\begin{list}{}{}
\item[\textbf{Phase estimation algorithm:}] Given are $W$,
$\ket{\psi'}$ (as a state of a quantum system) and the goal precision
$p$. An estimate of an eigenphase $\phi$ of $W$ is to be returned,
where the probability of $\phi$ is given by the population of
$\ket{\psi'}$ in the corresponding eigenspace.
\item[0.] Let $n$ be the smallest natural number such that
$2^n \geq 1/p$.
\begin{list}{}{}
\item[1.a.] 
Prepare $\kets{+}{a}$ in an ancilla qubit $\sysfnt{a}$ and
apply $\slb{\cntrl{W}}{aS}$ $2^{n-1}$ times. With the auxiliary assumption
that $\ket{\psi'}$ is an $e^{i\phi}$-eigenstate of $W$,
the effect is a phase kickback, changing $\kets{+}{a}$
to $(\kets{0}{a}+e^{i 2^{n-1}\phi}\kets{1}{a})/\sqrt{2}$.
\item[1.b.] Measure $\sysfnt{a}$ in the $\ket{+},\ket{-}$ basis, so that
measurement outcome $0$ ($1$) is associated with detecting $\ket{+}$
($\ket{-}$). Let $b'_n$ be the measurement
outcome. With the auxiliary assumption, the probability
that $b'_n=b_n$ is $\cos(\pi\epsilon/2)^2$.
\end{list}
\item[2] Do the following for each $k=(n-1),\ldots,1$:
\begin{list}{}{}
\item[2.a] Prepare $\kets{+}{a}$ in an ancilla qubit $\sysfnt{a}$ and
apply $\slb{\cntrl{W}}{aS}$ $2^{k-1}$ times. With the
auxiliary assumption, this changes $\kets{+}{a}$
to $(\kets{0}{a}+e^{i 2^{k-1}\phi}\kets{1}{a})/\sqrt{2}$.
\item[2.b] Compensate the phase of $\kets{1}{a}$ by changing it by
$e^{-i \pi[.b'_{k+1}\ldots b'_{n}]_2}$. With the auxiliary
assumption, this changes the state of the ancilla to
$(\kets{0}{a}+e^{i (2^{k-1}\phi-\pi[.b'_{k+1}\ldots
b'_{n}]_2)}\kets{1}{a})/\sqrt{2}$.
\item[2.c] Measure $\sysfnt{a}$  in the $\ket{+},\ket{-}$ basis
to obtain $b'_{k}$. With the auxiliary assumption and
if $b'_{l}=b_l$ for $l>k$, the probability that $b'_k=b_k$
is $\cos(\pi\epsilon/2^{n-k+1})^2$.
\end{list}
\item[3] Estimate $\phi$ as  $2\pi[.b'_1\ldots b'_n]_2$.
\end{list}
A step of the algorithm is depicted in Fig.~\ref{phase-est}.

The probability $P(\epsilon)$ that the value returned by the PEA is
$2\pi[.b_1\ldots b_n]_2$ is the product of the probabilities
$\cos(\pi\epsilon/2^{l})^2$ for $l=1,\ldots,n$ and is bounded below by
$\sin(\pi \epsilon)^2/(\pi \epsilon)^2$. This  bound can be obtained by
taking the limit $n \rightarrow \infty$ in $P(\epsilon)$. The worst
case is given for $|\epsilon|=1/2$, leading to the bound $P(\epsilon) \ge
4/\pi^2\approx 0.405$~\cite{cleve:qc1997b}. Since the goal precision is
$2^{-n}$, it is acceptable for the algorithm to obtain the next best
binary approximation to $\phi$. For this, the value obtained for $b_n'$
may not be the one with maximum probability, but the subsequent bits
$b_k'$ are always the best possible given $b_n'$. Taking this into
account, the probability that the phase returned is within $2^{-n}$ is
given by  $P(\epsilon)+P(1-\epsilon)\ge 8/\pi^2 \approx 0.81$
(see Note~\cite{note:fbound}).

The key step of the one-qubit phase estimation procedure is to modify
the phase kickback by the previously obtained phase estimate. This
differentiates it from an adaptive phase measurement method that
determines the bits of an approximation of $\phi/(2\pi)$ starting with
the most significant bit, and making sufficiently many measurements
with different phase compensations for each bit to achieve high
confidence level. This is the phase estimation method given
in~\cite{kitaev:qc1995a} and mentioned
in~\cite{giovannetti:qc2006a}, which approximates what is
done in practice for the efficient determination of an unknown
frequency or pulse time.

The resources required by the PEA, AEA and OEA
can be summarized as follows.
\begin{list}{}{}
\item[$\mathrm{PE}(W,\ket{\psi'},p)$:] This requires $N(p)=2^{\lceil
\log_2(1/p)\rceil}-1$ uses of $W$. $\ket{\psi'}$ is prepared once.
Here, $\lceil x\rceil$ denotes the least integer $m\geq x$.
\item[$\mathrm{AE}(U,\ket{\psi},p)$:] This calls $\mathrm{PE}$ once. It
requires $N(2p)$ uses of $S=VP_0V^\dagger UVP_0V^\dagger U^\dagger$ and
one use of $V$ to prepare the initial state. We count this as
being equivalent to $4N(2p)+1$ state preparations and $2N(2p)$
applications of $U$.
\item[$\mathrm{OE}(U,\ket{\psi},p)$:] This contains three calls to the
AEA with higher precision.  The total
resource count is $8 N(p/8)+4 N(p/2)+3$ state preparations and $4
N(p/8)+2 N(p/2)$ uses of $U$.
\end{list}
Since $N(p)$ is of order $1/p$, each of these algorithms uses
resources of order $1/p$.

\section{Confidence Bounds}
\label{sect:conf}

The PEA as described in the previous section obtains an estimate
$\phi_{\mathrm{est}}$ of an eigenphase $\phi$ such that the prior
probability that $\left|\phi_{\mathrm{est}}-\phi\right|< 2^{-n+1}\pi$
is at least $0.81$, regardless of the value of $\phi$, where
$n=\lceil\log_2(1/p)\rceil$. (The comparison of $\phi_{\mathrm{est}}$
to $\phi$ is modulo $2\pi$, so that
$\left|\phi_{\mathrm{est}}-\phi\right|$ is angular distance between
$e^{i\phi_{\mathrm{est}}}$ and $e^{i\phi}$.)  Thus, after having
obtained $\phi_{\mathrm{est}}$, we say that
$\phi=\phi_{\mathrm{est}}\pm 2^{-n+1}\pi$ with confidence level
$0.81$ or  $P[\phi_{\mathrm{est}}- 2^{-n+1}\pi< \phi < 
\phi_{\mathrm{est}}+ 2^{-n+1}\pi]=0.81$.  The error bound of
$2^{-n+1}\pi$ must not be confused with a standard deviation.  Suppose
that we use a single sample from a gaussian distribution with standard
deviation $\sigma$ to infer the mean.  We would expect that the
confidence level increases as $1-e^{-\Omega((\Delta/\sigma)^2)}$ for an error
bound of $\Delta$. (The notation $\Omega(x)$ means a quantity
asymptotically bounded below by something proportional to $x$, that is,
there exists a constant $C>0$ such that the quantity is eventually
bounded below by $Cx$.)  In general, it is desirable to have
confidence levels that increase at least exponentially as a function of
distance $\Delta$ or as a function of additional resources used. 
Unfortunately, for a single instance of the PEA, we cannot do better
than have confidence level $1-O(1/\Delta)$ for $\phi=\phi_{\mathrm{est}}\pm
2^{-n+1}\pi\Delta$~\cite{cleve:qc1997b}. (Here, $O(x)$ denotes a
quantity that is of order $x$, that is a quantity that is eventually
bounded above by $Cx$ for some constant $C$.  The meaning of
``eventually'' depends on context. Here it means ``for sufficiently
small $x$''. If the asymptotics of the argument require that it go to
infinity, it means ``for sufficiently large $x$''.)  The method
suggested in~\cite{cleve:qc1997b} for increasing the confidence level is to
use the PEA with a higher goal precision of $p/2^l$. However this
improves the confidence level on $\phi=\phi_{\mathrm{est}}\pm
2^{-n+1}\pi\Delta$ to only $1-\Omega(1/(\Delta 2^l))$ and requires a
$2^l$ resource overhead, which is not an efficient improvement in
confidence level.

A reasonable goal is to attain confidence level $c=1-e^{-\Omega(r)}$ that
$\phi=\phi_{\mathrm{est}}\pm 2^{-n+1}\pi$ with a resource overhead of a
factor of $O(r)$.  This modifies the resource counts from the previous
section from $O(1/p)$ to $O(|\log(1-c)|/p)$, where $c$ is the
confidence level achieved.  To attain this goal, we modify each step of the
PEA by including repetition to improve the confidence level that acceptable
values for the bits are determined.  Let the two nearest $n$-digit
binary approximations to $\phi/(2\pi)$ be given by
$\phi/(2\pi)=[.b_1\ldots b_n]_2+\delta/2^n$ and
$\phi/(2\pi)=[.\tilde{b}_1\ldots \tilde{b}_n]_2+(\delta-1)/2^n$, where
$0\leq\delta<1$.  We wish to obtain one of these approximations with
high confidence level.  For the first step of the PEA, we perform two sets 
of $r$ experiments to obtain a good estimate of
$\delta'=\pi(\delta+b_n)$. The first set consists of $r$
$(\kets{+}{a},\kets{-}{a})$-measurements of the state
$\cntrl{W}^{2^{n-1}}\kets{+}{a}\kets{\psi}{S}$. The second consists of $r$
$(\kets{+}{a},\kets{-}{a})$-measurements of the state
$\cntrl{W}^{2^{n-1}}(\kets{0}{a}-i\kets{1}{a})/\sqrt{2}\kets{\psi}{S}$. Let
$x_1,x_2$ be the sample means of the measurement outcomes of the two
sets of experiments. In the limit of large $r$, $x_1$ and $x_2$
approach $\sin(\delta'/2)^2$ and $\sin(\delta'/2-\pi/4)^2$,
respectively. We have
\begin{equation}
\sin(\delta') = \cos(\delta'-\pi/2)=1-2\sin(\delta'/2-\pi/4)^2,\;
\cos(\delta') = 1-2\sin(\delta'/2)^2,
\end{equation}
so we can estimate $\delta'$ from $x_1$ and $x_2$ by letting
$\delta'_{\mathrm{est}}$ be the phase of the complex vector $(1-2x_1) +
i(1-2x_2)$. The probability of the event $E$ that $\delta'$ differs
from $\delta'_{\mathrm{est}}$ by more than $\pi/4$ modulo $2\pi$ can be
bounded as follows.  For this event, $|\sin(\delta')+i\cos(\delta') -
((1-2x_1) + i(1-2x_2))|^2 \geq 1/2$.  It follows that either
$|\sin(\delta'/2)^2-x_1|\geq 1/4$ or $|\sin(\delta'/2-\pi/4)^2-x_2|\geq
1/4$. The probability of each of these possibilities is bounded by the
probability that the mean of $r$ samples of the binomial distribution
with probability $p$ of outcome $1$ differs from $p$ by at least
$x=1/4$. The probability of this event is bounded by $2e^{-2rx^2} =
2e^{-r/8}$ (Hoeffding's bound~\cite{hoeffding:qc1963a}). This bound can
now be doubled to obtain a bound of $4e^{-r/8}$ on the probability of
$E$.

Let $a_n=1$ if $\delta'_{\mathrm{est}}$ is closer to $\pi$ than $0$,
and $a_n=0$ otherwise. Then $a_n=b_n$ or $a_n=\tilde{b}_n$. Which
equality holds does not affect the subsequent arguments, so without
loss of generality, assume that $a_n=b_n$.  Suppose that event $E$ did
not happen and that we have correctly obtained
$a_n=b_n,\ldots,a_{k+1}=b_{k+1}$.  For the step of the algorithm that
determines the $k$'th bit, modify the original step by compensating the
phase of $\kets{1}{a}$ by $e^{-i(\pi[.b_{k+1}\ldots
b_{n-1}]_2+\delta'_{\mathrm{est}}/2^{n-k})}$ and repeating the
measurement $r$ times. We set $a_k=1$ if the majority of the
measurement outcomes is $1$ and $a_k=0$ otherwise.  For each
measurement, the probability that the measurement outcome does not
agree with $b_k$ is at most
$\sin((\delta'-\delta'_{\mathrm{est}})/2^{n-k+1})^2$.  Our assumptions
imply that this is at most $\sin(\pi/2^{n-k+3})^2\leq
(\pi/2^{n-k+3})^2$. Using Hoeffding's bound again, the probability that
$a_k\not=b_k$ is bounded by $2e^{-2r(1/2-(\pi/2^{n-k+3})^2)}<
2e^{-r/2}$ (for a loose upper bound).

Summing the probabilities, we find that the probability that we do
not learn $b_1\ldots b_n$ or $\tilde{b}_1\ldots \tilde{b}_n$ is bounded
by $x(n,r)=2(n-1) e^{-r/2}+4 e^{-r/8}$. We can therefore say that the
modified PEA yields the desired phase to within $\pi/2^{n-1}$ with
confidence level $1-x(n,r)$, where $x(n,r)$ decreases exponentially in $r$. 
Note again that this confidence bound still should not be confused with a
similar confidence bound for a gaussian random variable.  Increasing
the confidence bound does not result in the expected increase in
confidence level. In order to have confidence level increasing exponentially toward
$1$ with increasing confidence bound and an additional overhead of
at most $O(|\log(p)|)$, we can repeat the determination of the $k$'th
bit $2^{n-k}r$ instead of $r$ many times.

For the purpose of having high confidence level in the precision with which a
quantity is estimated, our algorithms require the confidence level goal as
an input. The modified PEA may be outlined
as follows.
\begin{list}{}{}
\item[\textbf{Modified Phase estimation algorithm:}] Given are $W$,
$\kets{\psi'}{S}$, a goal precision $p$ and a goal confidence level $c$. An
eigenphase $\phi$ of $W$ is to be returned, where the probability of
$\phi$ is given by the population of $\ket{\psi'}$ in the
corresponding eigenspace. The final state of $\sysfnt{S}$
consists of states with eigenphases in the range $\phi\pm p$
with prior probability at least $c$.
\item[0.] Let $n$ be the smallest natural number such that
$2^n \geq 1/p$. Let $r$ be the smallest natural number
such that $x(n,r)<(1-c)$.
\item[1.] Obtain $\delta'_{\mathrm{est}}$ with the two sets
of $r$ measurements described above. Let $a_n=1$
if $\delta'_{\mathrm{est}}$ is closer to $\pi$ than $0$ and $a_n=0$
otherwise.
\item[2] Do the following for each $k=(n-1),\ldots,1$, in this order:
\begin{list}{}{}
\item[2.a] Obtain an estimate of the $k$'th bit $a_k$ of a binary
approximation to $\phi/(2\pi)$ by $r$ repetitions of the measurement
of steps 2.a-c given previously, but with a phase compensation that
uses $\delta'_{\mathrm{est}}$ as well as the previously obtained bits.
\end{list}
\item[3] Return $2\pi[.a_1\ldots a_n]_2$.
\end{list}
We define $\mathrm{PE}(W,\ket{\psi'},p,c)$ to be the value returned by
the modified PEA.  

The resources required grow by a factor of less than $2r$, where
$r=O(|\log(1-c)|)$. The constant hidden by the order notation may be
determined from the expression for $r$ in step 0 and is not very
large.  To modify the AEA to attain confidence level $c$, it suffices to
change the call to $\mathrm{PE}$ by including $c$ as an argument. 
Because the OEA has three independent calls to the AEA, it needs to
make these calls with confidence level arguments of $1-(1-c)/3$ to ensure
that the final confidence level is $c$. The resource requirements of all
three algorithms are $O(|\log(1-c)|/p)$, where this applies to both the
uses of $U$ and of the state preparation operator $V$ in the case of
the AEA and OEA.

\section{Expectation Estimation}
\label{sect:exptobs}

Let $A$ be an observable and assume that it is possible to evolve under
$\pm A$ for any amount of time. This means that we can implement the
unitary operator $e^{-iAt}$ for any $t$.  The traditional idealized
procedure for measuring $\langle A\rangle = \trace(A\rho)$ is to adjoin
a system consisting of a quantum particle in one dimension with
momentum observable $\hat p$ and apply the coupled evolution
$e^{-iA\tensor\hat p}$ to the initial state $\rho\tensor\ketbra{0}{0}$,
where $\ket{0}$ is the position ``eigenstate'' with eigenvalue $0$.
Measuring the position of the particle yields a sample from the
distribution of eigenvalues of $A$~\cite{vonneumann:qc1932,
ortiz:qc2000a}. This procedure requires unbounded energy, both for
preparing $\ket{0}$ and to implement the coupled evolution. Performing
this measurement $N$ times yields an estimate of $\langle A\rangle$
with precision of order $\mathrm{var}(A)/\sqrt{N}$, where the variance
is $\mathrm{var}(A)=\langle (A-\langle A\rangle)^2\rangle$.  It is
desirable to improve the precision and to properly account for the
resources required to implement the coupling.

We focus on measurement methods that can be implemented in a quantum
information processor. In order to accomplish this, some prior
knowledge of the distribution of eigenvalues of $A$ with respect to
$\rho$ is required.  Suppose we have an upper bound $b$ on
$|\trace(A\rho)|$ and a bound on the tail distribution $F(\Delta)\geq
\trace([|A-\langle A\rangle|>\Delta]\rho)$, where $[|A-\langle
A\rangle|>\Delta]$ denotes the projection operator onto eigenspaces of
$A$ with eigenvalues $\lambda$ satisfying $|\lambda-\langle
A\rangle|>\Delta$. That is, $F(\Delta)\geq \sum_{|\lambda-\langle
A\rangle|>\Delta} p_\lambda$ with $p_\lambda=\trace
(\ket{\lambda}\bra{\lambda} \rho)$.  Without loss of generality, $F$
is non-increasing in $\Delta$.  An estimate on the tail distribution
is needed to guarantee the confidence bounds on $\trace(A\rho)$
derived from measurements by finite means.  Here are some examples: If
the maximum eigenvalue of $A$ is $\lambda_{\mathrm{max}}$, we can set
$b=\lambda_{\mathrm{max}}$ and use $F(\Delta) = 1$ if $\Delta
<\lambda_{\mathrm{max}}$ and $F(\Delta)=0$, otherwise.  Suppose that
we have an upper bound $v$ on the variance $\mathrm{var}(A)$. If we
know that the distribution of eigenvalues of $A$ is gaussian, we can
estimate $F(\Delta)$ by means of the error function for gaussian
distributions. With no such prior knowledge, the best estimate is
$F(\Delta)=\min(1,v/\Delta^2)$. (Observe that $v \geq
\Delta^2\sum_{|\lambda-\langle A\rangle|>\Delta} p_\lambda$.)  Such
``polynomial'' tails result in significant overheads for measuring
$\langle A\rangle$. ``Good'' tails should drop off at least
exponentially for large $\Delta$ (``exponential tails'').

We give an EEA based on overlap estimation.  The relevant resources
for the EEA are the number $M$ of times a unitary operator of the form
$e^{-iAt}$ is used, the total time $T$ that we evolve under $A$, and
the number $N$ of preparations of $\rho$. The total time $T$ is the
sum of the absolute values of exponents $t$ in uses of $e^{-iAt}$.
For applying the OEA, it is necessary to be able
to evolve under $-\!A$ as well as $A$.  If the evolution is implemented
by means of quantum networks, this poses no difficulty. However, if
the evolution uses physical Hamiltonians, this is a nontrivial
requirement.  The complexity of realizing $e^{-iAt}$ may depend on $t$
and the precision required. Since this is strongly dependent on $A$
and the methods used for evolving under $A$, we do not take this into
consideration and assume that the error in the implementation of
$e^{-iAt}$ is sufficiently small compared to the goal precision.  In
most cases of interest this is justified by results such as those
in~\cite{berry:qc2005a}, which show that for a large class of
operators $A$, $e^{-iAt}$ can be implemented with resources of order
$t^{1+\alpha'}/\epsilon^{\alpha'}$, where $\epsilon$ is the error of the
implementation and $\alpha'$ is arbitrarily small.

For exponential tails $F$, our algorithm achieves
$M,N=O(1/p^{1+\alpha})$ and $T=O(1/p)$ for arbitrarily small
$\alpha$. The order notation hides constants and an initialization
cost that depends on $b$ and $F$.  The strategy of the algorithm is to
measure $\trace(e^{-iAt}\rho)$ for various $t$. In the limit of small
$t$, $\trace(e^{-iAt}\rho) = 1+O(t^2)-i(\langle A\rangle t+O(t^3))$, so
that $\langle A \rangle$ can be determined to $O(t^3)$ from the
imaginary part of $\trace(e^{-iAt}\rho)$. The first problem is to make
an initial determination of $\langle A\rangle$ to within a deviation of
$A$ as determined by $F$. This is an issue when $b$ is large compared
to the deviation. To solve the first problem, we can use phase
estimation. We also give a more efficient method based on amplitude
estimation.  The second problem is to avoid excessive resources to achieve the
desired precision while making $t$ small.  To solve this problem
requires choosing $t$ carefully and taking advantage of  higher-order
approximations of $\langle A\rangle$ by linear combinations of
$\trace(e^{-iAt}\rho)$ for different times $t$.

To bound the systematic error in the approximation of $\langle
A\rangle$ by $i\trace(e^{-iAt}\rho)$, note that
$|\Im(e^{i\theta})-\theta|\leq\theta^3/6$. To see this it is
sufficient to bound the Lagrange remainder of the Taylor series of
$\sin(\theta)$.  This bound suffices for achieving $\alpha=1/2$ in the
bounds on $M$ and $N$. Reducing $\alpha$ requires a better
approximation, which we can derive from the Taylor series of the
principal branch of $\ln(x+1)$.  For $|x|<1$,
\begin{equation}
|\ln(x+1) -\sum_{k=1}^K(-1)^{k-1}x^k/k| \leq
  |x|^{K+1}/((K+1)(1-|x|)^{K+1}).
\end{equation}
To apply these series to the problem of approximating $\langle
A\rangle$, we compute
\begin{equation}
\label{eq:lgapp}
\sum_{k=1}^K(-1)^{k-1}(e^{-iBt}-1)^k/k
  = \sum_{l=0}^K C_l e^{-iBlt},
\end{equation}
for real constants $C_l$ satisfying $|C_l|\leq 2^K$. In particular, if
$B$ is an operator satisfying $|B|<x/t$, we can estimate
\begin{equation}
\label{eq:lgbapp}
\left|t\,\trace(B\rho)+
    \sum_{l=0}^K C_l \ \Im \ \trace(e^{-iBlt}\rho)
  \right| \leq
  |x|^{K+1}/((K+1)(1-|x|)^{K+1}).
\end{equation}

Define $G_e(\Delta) = \Delta F(\Delta)+\int_{\Delta}^\infty F(s)ds$.
Then $G_e(\Delta)$ is an upper bound on the contribution to the mean
from eigenvalues of $A$ that differ from the mean by more than
$\Delta$. That is, $G_e(\Delta) \geq \trace(|A-\langle
A\rangle|[|A-\langle A\rangle| > \Delta]\rho)= \sum_{|\lambda-\langle
A\rangle|>\Delta} |\lambda-\langle A\rangle| \ p_\lambda$.  Like
$F(\Delta)$, $G_e(\Delta)$ is non-increasing.  We assume that a
non-increasing bound $G(\Delta)\geq G_e(\Delta)$ is known and that
$G(\Delta)\rightarrow 0$ as $\Delta\rightarrow\infty$. Because
$F(\Delta) \leq G_e(\Delta)/\Delta$, we can use $G$ to bound both
$G_e$ and $F$.  For $x>0$, define $G^{-1}(x) = \inf\{\Delta|G(\Delta)
\leq x\}$.  The behavior of $G^{-1}$ as $x$ goes to $0$
determines the resource requirements for the EEA. If
$A$ is a bounded operator with bound $\lambda_{\mathrm{\max}}$, then
we can use $G^{-1}(x)\leq \lambda_{\mathrm{max}}$ independent of
$x>0$.  If $F$ is exponentially decaying, then so is $G$, and
$G^{-1}(x)=O(|\log(x)|)$.  For polynomial tails with $F(\Delta) =
O(1/\Delta^{2+\beta})$, we have $G(\Delta)=O(1/\Delta^{1+\beta})$ and
$G^{-1}(x) = O(1/x^{1/(1+\beta)})$.

The EEA has two stages.  The first is an initialization procedure to
determine $\langle A\rangle$ with an initial precision that is of the
order of a bound on the deviation of $A$ from its mean, where the
deviation is determined from $F$ and $G$. This initialization procedure
involves phase estimation to sample from the eigenvalue distribution of
$A$. Its purpose is to remove offsets in the case where the
expectation of $A$ may be very large compared to the width of the
distribution of eigenvalues as bounded by $F$ and $G$.  The second
stage zooms in on $\trace(A\rho)$ by use of the overlap estimation
procedure.  As before, we can assume without loss of generality that
$\rho$ is pure, $\rho=\ketbra{\psi}{\psi}$. We first give a version of
the EEA that achieves $M,N=O(1/p^{3/2})$ and then refine the algorithm
to achieve better asymptotic efficiency.
\begin{list}{}{}
\item[\textbf{Expectation estimation algorithm:}] Given are $A$,
$\ket{\psi}$ (in terms of a preparation unitary
$V:\ket{0}\mapsto\ket{\psi}$), a goal precision $p$ and the desired confidence level $c$. The
returned value is within $p$ of $\langle A\rangle =
\trace(A\ketbra{\psi}{\psi})$ with probability at least $c$.
\begin{list}{}{}
\item[Stage I.]
\item[0.] Choose $\Delta$ such that $F(\Delta/2)< 1/4$ and $\Delta
\geq p$.  $\Delta$ should be chosen as small as possible.  Let
$t_i=\pi/(4(b+\Delta))$.  Let $r$ be the minimum natural number such
that $2e^{-r/8}\leq(1-c)/4$ and set $c'$ according to the identity
$r(1-c')=(1-c)/4$.
\item[1.] Obtain $\Lambda_1,\ldots,\Lambda_r$ from $r$ instances of
the PEA, $\Lambda_k=\mathrm{PE}(e^{-iAt_i},\ket{\psi},\Delta
t_i/2,c')$, where $2\pi$ is subtracted for any return values between
$\pi$ and $2\pi$ to ensure that $-\pi\leq\Lambda_k<\pi$.
\item[2.] Let $\Lambda_m$ be the median of
$\Lambda_1,\ldots,\Lambda_r$. We show below that the probability that
$|\Lambda_m/t_i+\langle A\rangle|>\Delta$ is bounded by
$2e^{-r/8}+r(1-c')\leq(1-c)/2$.
\item[3.] Let $a_0=-\Lambda_m/t_i$. We expect $a_0$ to be within $\Delta$
of $\langle A\rangle$ with confidence level $1-(1-c)/2$.
\end{list}
\begin{list}{}{}
\item[Stage II.] If $p=\Delta$, return $a_0$ and skip this stage.
\item[0.] Choose $\theta_{\max}$ and $t$ so that they satisfy 
\begin{equation}
\begin{array}[b]{lrcl}
\textrm{(A)}\hspace*{1in}&
\theta_{\max}^3/6&\leq& (t/2)p/4,\\
\textrm{(B)}& G(\theta_{\max}/t)&\leq&
\theta_{\max}p/8,\\ 
\textrm{(C)}& \theta_{\max}&\leq& 1,\\
\textrm{(D)}& t\Delta&\leq&\theta_{\max}.\hspace*{1in} 
\end{array}
\label{eq:stgIIcnstr}
\end{equation}
The constraints and how they can be satisfied are explained below. The
parameter $t$ should be chosen as large as possible to minimize
resource requirements.
\item[1.] Obtain $x={\rm OE}(e^{-i(A-a_0)(t/2)},\ket{\psi},(t/2)p/4,1-(1-c)/2)$.
\item[2.] Return $-\Im(x)/(t/2)+a_0$.
\end{list}
\end{list}
Consider stage I of the algorithm. The probability that
$|\Lambda_m/t_i+\langle A\rangle|>\Delta$ may be bounded as follows.
The choice of $t_i$ ensures that eigenvalues $\Lambda$ of $-At_i$
within $\Delta t_i$ of the mean are between $\pm \pi/4$ and do not get
``aliased'' by $e^{-iAt_i}$ in the calls to the PEA. With probability 
at least $1-r(1-c')$, each $\Lambda_k$ returned by these calls is
within $t_i\Delta/2$ of an eigenvalue of $-At_i$ sampled
according to the probability distribution induced by $\ket{\psi}$. 
Assume that the event described in the previous sentence occurred.  The
probability that $|\Lambda_m/t_i+\langle A\rangle| > \Delta$ is upper
bounded by the probability that at least $\lceil r/2\rceil$ of the $r$
samples fall outside the range $[-\langle A\rangle t_i-\Delta
t_i,-\langle A\rangle t_i+\Delta t_i]$. The choice of $\Delta$ with
respect to $F$ implies that Hoeffding's bound can be applied to bound
this probability by $2 e^{-r/8}$.  Thus, we can bound the overall prior
probability $P$ that $|\Lambda_m/t_i+\langle A\rangle| > \Delta$ by
$P<2e^{-r/8}+r(1-c')\leq (1-c)/2$.

The resources required for stage I include $N=r=O(|\log(1-c)|)$
preparations of $\ket{\psi}$, $M=O(|\log(1-c)|(b+\Delta)/\Delta)$ uses
of $e^{-iAs}$ (specifically, $M$ is within a factor of $2$ of
$2r/\Delta t_i$) and a total evolution time of
$T=O(|\log(1-c)|/\Delta)$ (where $T$ is within a factor of $2$ of
$2r\Delta$).  Note that none of these resource bounds depend on the
$p$ and that $\Delta$ is a bound on a
deviation of $A$ from the mean with respect to $\ket{\psi}$. Also, if
$\Delta$ is of the same order as $b$, the formulation of stage I of the
algorithm is such that the uses of phase estimation require minimal
precision. In fact, in this case, stage I of the algorithm could be
skipped with minor adjustments to stage II. We show below that
stage I can be modified so that the overhead as a function of $b$ is
logarithmic. The modification requires that the number of state
preparations $N$ is of the same order as $M$.

In the special case of parameter estimation (see the introduction),
$\Delta=p$. Consequently stage II is skipped and the resources of
stage I are the total resources required. The algorithm therefore
achieves the optimal $O(1/p)$ resource requirements for this
situation.

Consider stage II of the algorithm. The error
$|-\Im(x)/(t/2)+a_0-\langle A\rangle|$ may be bounded as follows.  We
assume that all the precision constraints of stage I and II are
satisfied. The confidence level that this is true is $c$ overall.  With this
assumption, $x/(t/2)$ is within $p/4$ (the ``precision error'') of
$\trace(e^{-i(A-a_0)(t/2)}\rho)/(t/2)$.  There are three contributions
to the ``approximation error'', which is the difference between
$-\Im\,\trace(e^{-i(A-a_0)(t/2)}\rho)/(t/2)$ and $\trace((A-a_0)\rho)$.
For all contributions, we have to consider the fact that $a_0$
approximates $\langle A\rangle$ to within only $\Delta$, which is why
we need constraint (D) of Eq.~(\ref{eq:stgIIcnstr}).  The first arises
from eigenvalues of $(A-a_0)(t/2)$ in
$[-\theta_{\max},+\theta_{\max}]$ due to $|\Im(e^{i\theta})-\theta|$
not being zero and is bounded by $\theta_{\max}^3/(6(t/2))=p/4$
(constraint (A) of Eq.~(\ref{eq:stgIIcnstr})).  The second and third
come from eigenvalues of $(A-a_0)(t/2)$ outside
$[-\theta_{\max},+\theta_{\max}]$.  Constraint (D) of
Eq.~(\ref{eq:stgIIcnstr}) implies that $|(a_0-\langle A
\rangle)(t/2)|\leq \theta_{\max}/2$.  Constraints (B) and (C) of
Eq.~(\ref{eq:stgIIcnstr}) imply that the contribution to $\langle
A\rangle$ of eigenvalues differing from the mean by more than
$\theta_{\max}/(2(t/2))$ is at most $\theta_{\max}p/8\leq p/8$.
However the same eigenvalues still contribute to the measurement, each
contributing at most $1$ to $x$. Constraint (B) of
Eq.~(\ref{eq:stgIIcnstr}) together with the inequality $F(\Delta)\leq
G(\Delta)/\Delta$ imply that $F(\theta_{\max}/(2(t/2)))\leq tp/8$ so
this contribution has probability at most $tp/8$ and therefore adds at
most another $p/4$ (after dividing by $t/2$) to the approximation
error. Thus, the combination of the approximation and precision error
is less than $p$, as desired.  Clearly these estimates are suboptimal,
tighter choices of $\theta_{\max}$ and $t$ could be
made. However, this does not affect the asymptotics of the resource
requirements.

To find good solutions $\theta_{\max}$ and $t$ subject to the constraints given
in Eq.~(\ref{eq:stgIIcnstr}), we can rewrite the constraints as
follows:
\begin{equation}
\begin{array}[t]{lc}
\textrm{(A')}&
  G^{-1}(\theta_{\max}p/8)\leq \theta_{\max}/t \leq (p/8)/(\theta_{\max}^2/6),\\
\textrm{(B')}&
  \theta_{\max}\leq 1,\; \theta_{\max}/t \geq \Delta.
\end{array}
\label{eq:stgIIcnstr_b}
\end{equation}
The first inequality of (A') is implied by constraint (B) and the
second by constraint (A) of Eq.~(\ref{eq:stgIIcnstr}).
To satisfy these constraints, we first find $\theta_{\max}\leq 1$ as
large as possible so that 
\begin{equation}
\begin{array}[c]{lc}
\textrm{(A'')} & \Delta \leq G^{-1}(\theta_{\rm max}p/8) \leq
 (p/8)/(\theta_{\max}^2/6) ,
\end{array}
\label{eq:stgIIcnstr_c}
\end{equation}
and then set $t=\theta_{\max}/G^{-1}(\theta_{\max}p/8)$.  Consider the
three examples of bounded, exponential and polynomial tails.  For the
case of bounded tails, constraint (A'') of Eq.~(\ref{eq:stgIIcnstr_c})
can be solved by setting $\theta_{\max}$ according to
$\lambda_{\max}=(p/8)/(\theta_{\max}^2/6)$, so that $\theta_{\max} =
(3p/(4\lambda_{\max}))^{1/2}$.
The parameter $t$ is given by $\theta_{\max}/\lambda_{\max} =
(3p/4)^{1/2}/\lambda_{\max}^{3/2}=\Omega(p^{1/2})$.  For the case of
exponential tails, we can use $G^{-1}(x) = O(|\log(x)|)$ to show that
$\theta_{\max}=\Omega((p/|\log(p)|)^{1/2})$ and
$t=\Omega(p^{1/2}/|\log(p)|^{3/2})$ (see Note \cite{note3}).  For polynomial tails
with $G^{-1}(x) = O(x^{-1/(1+\beta)})$, we get
$\theta_{\max}=\Omega(p^{(2+\beta)/(1+2\beta)})$ and
$t=\Omega(p^{(5+6\beta+\beta^2)/((1+\beta)(1+2\beta))})$ (see Note \cite{note4}).

The resource requirements for stage II of the EEA can be estimated as
$M=O(|\log(1-c)|/tp)$ uses of an exponential of the form $e^{-iAs}$,
$N=O(|\log(1-c)|/tp)$ state preparations, and a total time of
$T=O(|\log(1-c)|/p)$, in terms of the parameter $t$ computed in step 0
(of stage II). The dependence on $G$ shows up in the value of $t$.
With $t$ as computed in the previous paragraph, for bounded $A$, $M$
and $N$ are $O(|\log(1-c)|/p^{3/2})$. For exponential tails, $M$ and
$N$ are $O(|\log(1-c)|/(p/|\log(p)|)^{3/2})$.  For polynomial tails,
they are $O(|\log(1-c)|/p^{\gamma(\beta)})$, where $\gamma(\beta)$ is a
polynomial satisfying $\gamma(\beta) \rightarrow 1 + 1/2$ for
$\beta\rightarrow \infty$ and $\gamma(\beta) = 1 + 5$ for $\beta=0$.

To reduce the resource requirements of stage II of the EEA, we use
overlap estimation at multiple values of $t$ and
Eq.~(\ref{eq:lgapp}). Here is the modified stage. We assume that $K\geq
2$.
\begin{list}{}{}
\item[Stage II'.]
\item[0.] Choose $\theta_{\max}$ and $t$ so that they satisfy 
\begin{equation}
\begin{array}[b]{lrcl}
\textrm{(A)}\hspace*{1in}&
\theta_{\max}^{K+1}/((K+1)(1-\theta_{\max})^{K+1})&\leq& (t/2)p/4,\\
\textrm{(B)}& G(\theta_{\max}/t)&\leq&
\theta_{\max}p/(8K2^K),\\ 
\textrm{(C)}& \theta_{\max}&\leq& 1,\\
\textrm{(D)}& t\Delta&\leq&\theta_{\max}.\hspace*{1in} 
\end{array}
\label{eq:stgIIpcnstr}
\end{equation}
The parameter $t$ should be chosen as large as possible to minimize
resource requirements.
\item[1.] For $l=1,\ldots,K$, obtain $y_l={\rm
OE}(e^{-i(A-a_0)(lt/2)},\ket{\psi},(t/2)p/(4 K 2^K),1-(1-c)/(2K))$. Let
$y_0=1$.
\item[2.] Return $-\Im(\sum_{l=0}^K C_l y_l)/(t/2)+a_0$.
\end{list}
The precisions and the confidence levels in the calls to the OEA have been
adjusted so that the final answer has the correct precision and
confidence level. The explanation for this is similar to
that for the original stage II (see Note \cite{note5}). 

The earlier method for finding $\theta_{\max}$ and $t$ is readily
adapted to the constraints in stage II'. Constraint A'' of
Eq.~(\ref{eq:stgIIcnstr_c}) now reads as
\begin{equation}
\begin{array}[c]{lc}
\textrm{(A'')} & \Delta \leq G^{^-1}(\theta_{\max}p/(8K2^K)) \leq
(p/8)(K+1)(1-\theta_{\max})^{K+1}/\theta_{\max}^{K},
\end{array}
\label{eq:stgIIpcnstr_c}
\end{equation}
and we can set $t=\theta_{\max}/G^{-1}(\theta_{\max}p/(8K2^K))$.  To
simplify the right hand side of Eq.~(\ref{eq:stgIIpcnstr_c}), we add
the inequality $\theta_{\max}\leq 1/(K+1)$, and use the inequality
$1/4\leq (2/3)^3\leq (1-1/(K+1))^{K+1}$ (for $K\geq 2$) to replace the
right hand side by $(p/32)(K+1)/\theta_{\max}^K$.  Thus for bounded
tails, $\theta_{\max} = \Omega(\min(1/K,(Kp)^{1/K}))$ and
$t=\Omega(\min(1/K, (Kp)^{1/K}))$, where we give the asymptotic
dependence on $K$ explicitly but suppress parameters not depending on
$K$ or $p$ (see Note \cite{note6}). For exponential tails, $\theta_{\max} =
\Omega(\min(1/K,(p/|\log(p)|)^{1/K}))$ and
$t=\Omega(\min(1/(K(|\log(p)|+K)),p^{1/K}/(|\log(p)|^{1/K}(|\log(p)|+K))))$
(see Note \cite{note7}). For polynomial tails with exponent $\beta$,
$\theta_{\max} = \Omega(\min(1/K,p^{(2+\beta)/(K-1+K\beta)}))$ and
$t=\Omega(\min(1/(K(2^{K/(1+\beta)}(Kp^{-1})^{1/(1+\beta)})),
2^{-K/(1+\beta)}p^{(2+\beta)^2/((1+\beta)(K-1+K\beta))+1/(1+\beta)}))$
(see Note \cite{note8}).

With the expressions from the previous paragraph, we can estimate the
resources requirements of stage II'.  In terms of $t$, $M$ and $N$ are
$O(K^22^K|\log(1-c)|/tp)$,  and $T= O(K^32^K|\log(1-c)|/p)$, where the
powers of $K$ account for the $K$ calls to the OEA, the coefficient in 
the denominator of the precision, and in the case of $T$, the factor of
$l$ in the evolution time.  For bounded tails, we obtain
$M,N=O(|\log((1-c))|K^3
2^K/p^{1+1/(K)})$, where we have loosely
increased the power of $K$ by $1$ to account for the upper bound of
$O(1/K)$ on $t$.  For exponential tails, $M,N=O(|\log((1-c)/K)|K^4
2^K/(p/|\log(p)|)^{1+1/(K+1)})$ (with appropriate increases in the
power of $K$), and for polynomial tails, $M,N=O(|\log((1-c)/K)|K^4
2^{3K}/p^{\gamma(\beta,K)})$ (with conservative increases in the power
of $K$ and the exponent of $2$), where $\gamma(\beta,K)$ approaches
$1+1/(1+\beta)$ for large $K$.
Note that for $\beta=0$, this approaches
the ``classical'' resource bound as a function of precision.

\draftcomment{Is the bad behavior as a function of $\beta$ for
polynomial tails removable by simply continuing phase estimation? 
Or at any rate, might it be preferable in some regimes to just
use phase estimation? It might be worth having a look,
though note that we can always beat the $1/p^2$ bound
for $\beta>0$, albeit at a significant cost.
How does this compare to the one-dimensional particle based
method, taking the true energy cost into account in terms
of the energy's standard deviation?}

\draftcomment{Can uses of conditional $e^{-iAs}$ be eliminated?  This
requires modifying stage I and the OEA so
that the conditionals there are not used to determine the phase of the
overlap. That does not appear to be possible for the purpose used
here, that is, there is no obvious way to convert the expectation to
the norm of an amplitude without this trick.}

The final task of this section is to modify stage I so that the
dependence of the resource requirements on $b$ is logarithmic rather
than linear in $b$. The basic idea is to use logarithmic search to
reduce the uncertainty in $\langle A\rangle$ to $\Delta$.  Define $q$
by $b=q\Delta$.
\begin{list}{}{}
\item[Stage I'.]
\item[0.] Chose $\Delta$ minimal so that $G(\Delta)<\Delta/6$ and
$F(\Delta)<1/18$. Set the initial estimate of $\langle A\rangle$
to $a=0$ and the initial precision to $p_a=b=q\Delta$.
\item[1.] Repeat the following until  $p_a\leq \Delta$:
\item[1.a.] Set $t=1/(p_a+\Delta)$ and obtain
$x = \mathrm{OE}(e^{-i(A-a)t},\ket{\psi},1/18,1-(1-c)/(2\lceil\log_2(q)\rceil))$.
\item[1.b.] Update $a$ and $p_a$ according to the assignments
$a\leftarrow a-\Im(x)/t$ and 
$p_a\leftarrow (\Delta/6+(5/18)(p_a+\Delta)$.
\end{list}
We claim that at the end of this stage, we have determined $\langle
A\rangle$ to within $\Delta$ with overall confidence level $1-(1-c)/2$, so
that we can continue with the second stage, as before.  To verify the
claim, it is necessary to confirm that at the end of step 1.b., the
updated estimate $a$ of $\langle A\rangle$ has precision $p_a$. The
error in $a$ can be bounded as we have done for stage II.  Let $a_0$ be
the estimate of $A$ used in the call to the OEA. There is an error of
less than $1/(18t)=(p_{a_0}+\Delta)/18$ due to precision of $x$ in the
call to the OEA.  The remaining error is due to the approximation of
$\trace((A-a_0)t \rho)$ by $-\Im(\trace(e^{-i(A-a)t}\rho))$. For
eigenvalues $\lambda$ of $A$ within $1/t$ of $a$, this
is bounded by $|\lambda
t+\Im(e^{-i\lambda t})|\leq 1/6$, which translates into an
approximation error of at most $1/(6t) = (p_{a_0}+\Delta)/6$. 
Eigenvalues of $A$ further from $a$ than $1/t = p_{a_0}+\Delta$ are at
least $\Delta$ from $\langle A\rangle$. This requires the inductive
assumption that the $|a_0-\langle A\rangle|\leq p_{a_0}$. The
contribution to the mean from such eigenvalues is bounded by
$\Delta/6$, and the bias resulting from their contribution to $x$ is at
most $F(\Delta)/t=(p_{a_0}+\Delta)/18$.  Adding up the errors gives the
$p_a$ computed in step 1.b.  The confidence levels in the calls to the OEA
are chosen so that the final confidence level is $1-(1-c)/2$.  To see this
requires verifying that the number of calls of the OEA is at most
$\lceil\log_2(q)\rceil$. It suffices to show that if $p_{a_0}\geq
2\Delta$, then $\Delta/6+(5/18)(\Delta+p_{a_0})\leq p_{a_0}/2$. Rewrite
the left hand side as $(8/18)\Delta+(5/18)p_{a_0}$, which for
$p_{a_0}\geq 2\Delta$ is less than
$(4/18)p_{a_0}+(5/18)p_{a_0}=p_{a_0}/2$.

Each call to the OEA in stage II' has constant precision, which implies
that $M$ and $N$ are both $O(\log(q))=O(\log(b/\Delta))$ for large
$q$.  The total time $T$ is $O(1/\Delta)$.

\section{Parallelizability}
\label{sect:parallel}

To what extent are the algorithms given in the previous sections
parallelizable?   Consider the OEA.  At its core is the PEA with a
unitary operator $S$ that has two eigenvalues $e^{\pm i\phi}$ on the
relevant state space. In the sequential implementation, one of the
eigenvalues is eventually obtained with the desired precision. Which
eigenvalue is returned cannot be predicted beforehand. The initial
state is such that each one has equal probability. If it is possible to
deterministically (or near-deterministically) prepare an eigenstate
$\ket{\psi_{\phi}}$ with (say) eigenvalue $e^{i\phi}$ using
sufficiently few resources, then we can use the entanglement trick
in~\cite{bollinger:qc1996a} to parallelize the algorithm.  Instead of
applying $V$ sequentially $2^{k-1}$ many times to determine bit $k$ of
the phase, we prepare the entangled state
$(\kets{0\ldots0}{a}+\kets{1\ldots1}{a})/\sqrt{2}$ on $2^{k-1}$ ancilla
qubits and $2^{k-1}$ copies of $\ket{\psi_{\phi}}$. We next apply $\cntrl{S}$
between the $j$'th ancilla and the $j$'th copy of $\ket{\psi_{\phi}}$
and then make a measurement of
$(\kets{0\ldots0}{a}\pm\kets{1\ldots1}{a})/\sqrt{2})$. On a quantum
computer, the measurement requires decoding the superposition into a
qubit, which can be done with $O(2^k)$ gates.  The decoding procedure
can be parallelized to reduce the time to $O(k)$ (see Note \cite{note9}). Using
this trick reduces the time of the PEA to $O(\log(1/p))$ (the number of
bits to be determined), counting only the sequential uses of $U$ and
ignoring the complexity of preparing the initial states
$\ket{\psi_{\phi}}$ and the decoding overhead in the measurement. The
repetitions required for achieving the desired confidence level are trivially
parallelizable and do not contribute to the time.  It is possible to
reduce the time from $O(\log(1/p))$ to $O(1)$ by avoiding the
feed-forward phase correction used in the algorithm and reverting to
the algorithm in~\cite{kitaev:qc1995a} and mentioned
in~\cite{giovannetti:qc2006a}.
\begin{figure}
\includegraphics*[width=5.5in]{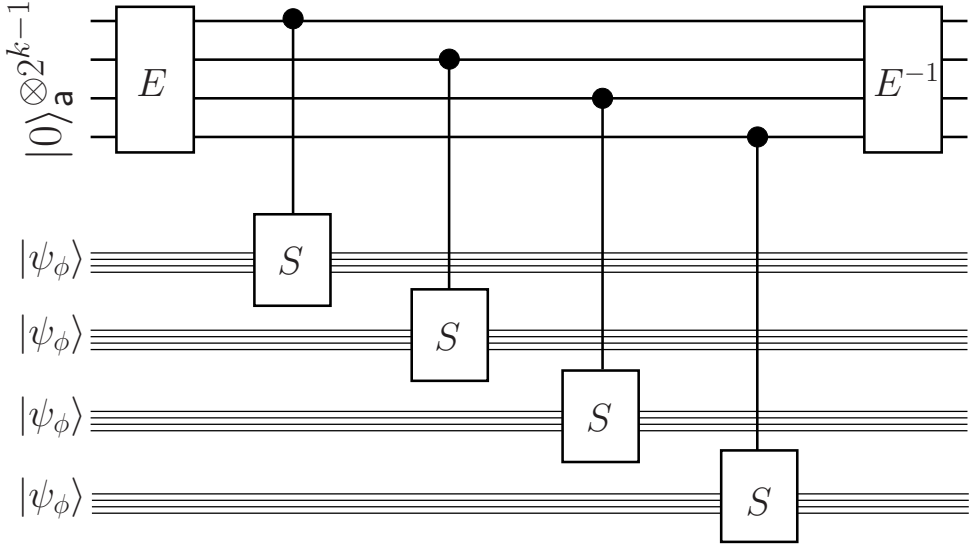}
\caption{Parallelization of the PEA algorithm to estimate the bit $k=3$
of the phase. This replaces the outlined parts of the network
in Fig.~\ref{phase-est}. 
$E$ is an entangler such that $E |0\rangle_{\sf
a}^{\otimes 2^{k-1}}= (\ket{0\cdots 0}_{\sf a}+\ket{1\cdots 1}_{\sf
a})/\sqrt{2}$, and $E |100 \cdots 0\rangle_{\sf a}= (\ket{0\cdots
0}_{\sf a}-\ket{1\cdots 1}_{\sf a})/\sqrt{2}$. $E^{-1}$ is the decoding
operation that maps $E^{-1}\ket{0\cdots 0}_{\sf a}= \ket{+ 0\cdots
0}_{\sf a}$, and $E^{-1}\ket{1\cdots 1}_{\sf a}= \ket{- 0\cdots
0}_{\sf a}$, where $\ket{\pm}=(\ket{0}\pm\ket{1})/\sqrt{2}$. 
The $k$'th bit is estimated from the measurement outcome of the first
ancilla qubit in the logical basis. }
\label{fig4}
\end{figure}

Based on the discussion in the previous paragraph, the main obstacle to
parallelizing the OEA is the preparation of $\ket{\psi_{\phi}}$. If
$\phi=2\arccos(|\trace(S\rho)|)$ is not close to $0$,
$\ket{\psi_{\phi}}$ can be prepared near deterministically with
relatively few resources as follows. Suppose we have a lower bound
$\epsilon$ on $\phi$. With the original initial state, use sequential
phase estimation with precision $\epsilon/2$ and confidence level
$1-(1-c)p/B$ to determine whether we have projected onto the eigenstate
$\ket{\psi_{\phi}}$ with eigenvalue $e^{i\phi}$ or the one with
$e^{-i\phi}$.  The occurrence of $p$ in the confidence level accounts for the
total number of states that need to be prepared.  The parameter $B$ is
a constant that provides an additional adjustment to the confidence level. 
It must be chosen sufficiently large, and other confidence level parameters
must be adjusted accordingly, to achieve the desired overall
confidence level.  If we have projected onto $\ket{\psi_{\phi}}$, return the
state. If not, either try again, or adapt the parallel PEA to use the
inverse operator $S^\dagger$ instead of $S$ for this instance of the
initial state. The (sequential) resources required are of the order of
$|\log((1-c)p)|/\epsilon$, but all the needed states can be prepared
in parallel. For $\epsilon$ constant, the time required by the parallel
PEA is increased by a factor of $O(|\log((1-c)p)|)$.  The parallel overlap
estimation for a unitary operator $U$ based on these variations of
phase estimation thus requires $O(|\log((1-c)p)|)$ time, provided
$|\bra{\psi}U\ket{\psi}|$ is not too close to $1$.

For $|\bra{\psi}U\ket{\psi}|$ close to $1$, the OEA is intrinsically
not parallelizable without increasing the total resource cost by a
factor of up to $O(\sqrt{p})$.  This is due to the results
in~\cite{zalka:qc1997a}, where it is shown that Grover's algorithm
cannot be parallelized without reducing the performance to that of
classical search.  For example, consider the problem of determining
which unique state $\ket{k}$ of the states $\ket{0},\ldots\ket{2^n}$
has its sign flipped by a ``black-box'' unitary operator $V$. This can
be done with $n$ many uses of the OEA by preparing the states
$\ket{\psi_b}$ that are uniform superpositions of the $\ket{i}$ for
which the number $i$ has $1$ as its $b$'th bit.  If
$\bra{\psi_b}U\ket{\psi_b}=1/2^{n-1}$, then the $b$'th bit of $k$ is
$1$.  If $\bra{\psi_b}U\ket{\psi_b}=0$, then it is $0$. It suffices to
use an unparametrized (Fig.~\ref{lift}) precision of $1/2^{n-1}$ and
confidence level sufficiently much bigger than $1-1/n$.  Because
$|1-\cos(\phi)|=O(\phi^2)$, the parameterized precision required is
$\Theta(1/2^{n/2})$.  ($\Theta(x)$ is a quantity that is both $O(x)$
and $\Omega(x)$.)  Thus $O(n 2^{n/2})$ sequential resources suffice,
which is close to the optimum attained by Grover's algorithm.
However, the results of~\cite{zalka:qc1997a} imply that implementing
quantum search with depth (sequential time) $d$ requires
$\Omega(2^{n}/d)$ uses of $V$ for $d < 2^{n/2}$.  This implies that to
achieve a parameterized precision of $\Theta(1/2^{n/2})$ for
$1-|\bra{\psi}U\ket{\psi}|= O(1/2^n)$ using time $O(2^{n/2}/P)$
requires $\Omega(2^{n/2}P)$ resources.

The EEA was described so that overlap estimation is used with small
$\phi$, and therefore can not be immediately parallelized without loss
of precision or larger resource requirements.  However, for the
version of overlap estimation needed for stages I' and II', it is only
the imaginary part of the overlap that is needed, and the parameters
are chosen so that the overlap's phase is expected to be within $1$ of
$0$ (because $\theta_{\max}\leq 1$). The actual precision required is
absolute in the overlap, not the parameterization of the overlap in
terms of the upper hemisphere in Fig.~\ref{lift}. This implies that we
can call the parallel overlap algorithm with an intentionally
suppressed overlap. If the desired overlap is $\bra{\psi}U\ket{\psi}$,
one way to suppress it is to replace $\slb{U}{S}$ by
$\slb{\cntrl{U}}{aS}$ and the initial state by
$(\slb{\one}{a}/2)\ketbras{\psi}{\psi}{S}$.  The suppression ensures
that the phases in the calls to the PEA are sufficiently
distinguishable to allow the near deterministic preparation of the
appropriate eigenstates discussed above. This adds at most a constant
overhead to the EEA due to the additional precision required to
account for the scaling associated with the overlap suppression.

\begin{acknowledgments}
We thank Ryan Epstein and Scott Glancy for their help in reviewing and
editing this manuscript. Contributions to this work by NIST, an
agency of the US government, are not subject to copyright laws. This
work was carried out under the auspices of the National Nuclear
Security Administration of the U.S. Department of Energy at Los Alamos
National Laboratory under Contract No. DE-AC52-06NA25396.
\end{acknowledgments}

\bibliography{journalDefs,obsexp}

\begin{thebibliography}{29}
\bibitem[{not({\natexlab{a}})}]{note1}
\bibinfo{note}{For the moment, we identify the precision with the maximum
  error. The correct formulation of precision in terms of confidences is given
  later. Because $|\bra{\psi}U\ket{\psi}|\leq 1$, the error in the value of
  $a^2$ is at most $p/2$. Similarly, the errors in $4b_0^2$ and $4b_{\pi/2}^2$
  are at most $p/2$. Here we used the fact that the parametrized precision
  returned by the AEA is also an upper bound on the precision of the
  (unparameterized) value returned. Thus $\Re(y)$ and $\Im(y)$ are both off by
  at most $p/2$. This error affects only the component of the returned value
  perpendicular to $\bra{\psi}U\ket{\psi}$ in the complex plane, to which it
  contributes at most $p/\sqrt{2}$. When these errors are lifted back to the
  upper unit hemisphere, they are bounded by $\sqrt{(p/2)^2+(p/\sqrt{2})^2} =
  p\sqrt{3}/2$, taking into account that the precision in $a$ is actually with
  respect to the parameterization. Our choice of precisions in the calls to
  amplitude estimation can be improved.}
\bibitem[{not({\natexlab{b}})}]{note:fbound}
\bibinfo{note}{Using $n$ as an explicit parameter, we have $P_0(\epsilon)=1$,
  and $P_n(\epsilon) = P_{n-1}(\epsilon/2)\cos(\pi\epsilon/2)^2$. The
  probability $P$ that the phase returned is within $2^{-n}$ is the probability
  $P_n(\epsilon)$ of getting the best approximation, plus the the probability
  $(1-\cos(\pi\epsilon/2)^2)=\sin(\pi\epsilon/2)^2=\cos(\pi(1-\epsilon)/2)^2$
  that the most significant bit is wrong times the probability
  $P_{n-1}((1-\epsilon)/2)$ that all subsequent bits are best possible. Thus
  $P=P_{n}(\epsilon)+ P_{n-1}((1-\epsilon)/2)\cos(\pi(1-\epsilon/2))^2 =
  P_{n}(\epsilon)+P_{n}(1-\epsilon)\geq 0.81$.}
\bibitem[{not({\natexlab{c}})}]{note3}
\bibinfo{note}{Let $C$ be a constant so that $G^{-1}(x) \leq C|\log(x)|$ for
  $x$ small enough. Since we are considering the asymptotic behavior of
  resources for small $p$, $x$ is small and the first inequality in constraint
  (A'') of Eq.~(\ref{eq:stgIIcnstr_c}) may be assumed to be satisfied. To solve
  constraint (A''), maximize $\theta_{\max}$ subject to
  $C|\log(\theta_{\max}p/8)| \leq (p/8)/(\theta^2_{\max}/6)$. Rewrite this
  inequality as $|\log(\theta_{\max}p/8)|\theta^2_{\max} \leq (3p/4)/C$. For
  $p$ small enough (and since $\theta_{\max} \leq 1)$, this is the same as
  $(|\log(\theta_{\max})|+|\log(p/8)|)\theta^2_{\max}\leq (3p/4)/C$. We can
  impose the additional constraint that $|\log(\theta_{\max})|\leq |\log(p/8)|$
  to write the inequality in the form $\theta^2_{\max}\leq
  ((3p/4)/C)/|\log(p/8)|$ or $\theta_{\max}\leq D'(p/|\log(p)|)^{1/2}$ for some
  constant $D'$. We can try a solution $\theta_{\max} = D''(p/|\log(p)|)^{1/2}
  = \Omega((p/|\log(p)|)^{1/2})$. For this solution, $|\log(\theta_{\max})|
  \leq |\log(D'')|+|\log(p)|/2 + \log(|\log(p)|)/2 \leq |\log(p/8)|$ for
  sufficiently small $p$. Hence the additional constraint is asymptotically
  satisfied.}
\bibitem[{not({\natexlab{d}})}]{note4}
\bibinfo{note}{Let $C$ be a constant so that $G^{-1}(x)\leq C x^{-1/(1+\beta)}$
  for $x$ small enough. As in~\cite{note3}, we are interested in the behavior
  for small $p$. To solve constraint (A''), maximize $\theta_{\max}$ subject to
  $C (\theta_{\max}p/8)^{-1/(1+\beta)}\leq (p/8)/(\theta_{\max}^2/6)$.
  Equivalently, $\theta_{\max}^{2-1/(1+\beta)} =
  \theta_{\max}^{(1+2\beta)/(1+\beta)}\leq D'p^{1+1/(1+\beta)} = D'
  p^{(2+\beta)/(1+\beta)}$ for some constant $D'$. Hence $\theta_{\max} =
  \Omega(p^{(2+\beta)/(1+2\beta)})$ works. From this we get
  $t=\theta_{\max}/G^{-1}(\theta_{\max}p/8) = \Omega(p^{(2+\beta)/(1+2\beta)}
  (p^{(2+\beta)/(1+2\beta)+1})^{1/(1+\beta)}) =
  \Omega(p^{(5+6\beta+\beta^2)/((1+2\beta)(1+\beta))})$.}
\bibitem[{not({\natexlab{e}})}]{note5}
\bibinfo{note}{The prior probability that all the $y_l$ are within
  $(t/2)p/(4K2^K)$ of the true overlap in the calls to the OEA is at least
  $(1-K (1-c)/(2K))=1-(1-c)/2$. Thus the confidence for stage II' matches that
  of stage II. Assuming that all the $y_l$ have the stated precision, the
  difference $|\sum_{l=0}^KC_l y_l-\sum_{l=0}^K C_l
  \trace(e^{-i(A-a_0)(lt/2)}\rho)|$ is bounded by $B=\sum_{l=0}^KC_l
  (t/2)p/(4K2^K)$. Since $|C_l|\leq 2^K$, $B\leq (t/2)p/4$. As before, the
  approximation error
  $|-\Im(\sum_{l=0}^KC_l\trace(e^{-i(A-a_0)(lt/2)}\rho))/(t/2)
  -\trace((A-a_0)\rho)|$ has three contributions. The first is due to the error
  term in Eq.~(\ref{eq:lgbapp}) for eigenvalues of $(A-a_0)(t/2)$ in
  $[-\theta_{\max},+\theta_{\max}]$. This is bounded by
  $\theta_{\max}^{K+1}/((K+1)(1-\theta_{\max})^{K+1}(t/2))$. Constraint (A)
  implies that it is at most $p/4$. The second and third contributions are due
  to other eigenvalues of $(A-a_0)(t/2)$. The contribution to the mean of these
  eigenvalues can be bounded by using constraints (B), (C) and (D). These
  eigenvalues differ from $a_0$ by at least $ 2\theta_{\max}/t$. According to
  constraint (D) and the correctness of stage I, they therefore differ from the
  mean by at least $\theta_{\max}/t$. By use of (B) and (C), their contribution
  to the mean is bounded by $p/(8K2^K)$. Each such eigenvalue still contributes
  to the values returned by the calls to the OEA, changing each $y_l$ by at
  most $F(\theta_{\max}/t)\leq tp/(8K2^K)$, which changes the returned value by
  at most $p/4$.}
\bibitem[{not({\natexlab{f}})}]{note6}
\bibinfo{note}{In this case, $G^{-1}(x)\leq \lambda_{\max}$. Thus we need
  $\theta_{\max}^K\leq (p/32)(K+1)/\lambda_{\max}$ and can set
  $t=\theta_{\max}/\lambda_{\max}$. Asymptotically, $K+1$ is equivalent to $K$
  in the expressions obtained.}
\bibitem[{not({\natexlab{g}})}]{note7}
\bibinfo{note}{Use $G^{-1}(x)\leq C|\log(x)|$ for sufficiently small $x$. We
  therefore need $C|\log(\theta_{\max}p/(8K2^K))| \leq
  (p/32)(K+1)/\theta_{\max}^K$. Because $\log(8K2^K) \leq 3K$, it is sufficient
  to satisfy $(|\log(\theta_{\max})|+|\log(p)|+3K)\theta_{\max}^K\leq Kp/32$.
  Add the additional constraint $|\log(\theta_{\max})|\leq |\log(p)|$, so that
  $\theta_{\max}^K \leq Kp/(|\log(p)|+3K)$ suffices. We can therefore set
  $\theta_{\max} = (Kp/(|\log(p)|+3K))^{1/K}=\Omega((p/|\log(p)|)^{1/K})$.
  Observe that $|\log(\theta_{\max})|$ satisfies the additional constraint for
  sufficiently small $p$, independent of $K$. To obtain $t$, note that
  $G^{-1}(\theta_{\max}p/(8K2^K)) = O(|\log(p)|+K)$, where we used the order
  notation to absorb constants.}
\bibitem[{not({\natexlab{h}})}]{note8}
\bibinfo{note}{Here $G^{-1}(x)\leq C x^{-1/(1+\beta)}$ for sufficiently small
  $x$, so we solve $(\theta_{\max}p/(8K2^K))^{-1/(1+\beta)} \leq
  (p/32)(K+1)/\theta_{\max}^K$. It is sufficient to solve
  $\theta_{\max}^{K-1/(1+\beta)} \leq D2^{-K/(1+\beta)}p^{1+1/(1+\beta)}$ for
  some sufficiently small constant $D$. Here we used the fact that
  $(K+1)/K^{1/(1+\beta)}= \Omega(1)$. Thus we can set $\theta_{\max} =
  \Omega(2^{-K/(K-1+K\beta)}p^{(2+\beta)/(K-1+K\beta)})$. The first factor is
  $\Omega(1)$. To bound $t$, $G^{-1}(\theta_{\max}p/(8K2^K)) =
  O(2^{K/(1+\beta)}p^{-(2+\beta)/((1+\beta)(K-1+K\beta))-1/(1+\beta)})$. Thus
  $t =
  \Omega(2^{-K/(1+\beta)}p^{(2+\beta)^2/((1+\beta)(K-1+K\beta))+1/(1+\beta)})$%
.}
\bibitem[{not({\natexlab{i}})}]{note9}
\bibinfo{note}{It suffices to assign the qubits to the leaves of a binary tree.
  The decoding proceeds recursively by applying CNOTs to pairs of leaves with a
  common parent, removing the target qubit, assigning the control qubit to the
  parent and removing the leaves from the tree. The qubit that ends up at the
  root of the tree is measured in the $\ket{+},\ket{-}$ basis.}
\end{thebibliography}

\end{document}